\documentclass{aa}
\usepackage{amsmath}
\usepackage[varg]{txfonts}
\usepackage{graphicx,amssymb,float,subfigure}
\usepackage{natbib}
\bibpunct{(}{)}{;}{a}{}{,}
\newcommand{\ideal}[1]{\ensuremath{\overset{\infty}{#1}}}

\newcommand{\dd}{\ensuremath{\mathrm{d}}}
\newcommand{\DD}{\ensuremath{\mathrm{D}}}

\begin{document}

\title{A localised subgrid scale model for fluid dynamical simulations in astrophysics I:
    Theory and numerical tests}

\author{W. Schmidt}
\author{W. Schmidt\inst{1,2} \and J. C. Niemeyer\inst{1} \and W. Hillebrandt\inst{2}}

\institute{Lehrstuhl f\"{u}r Astronomie, Institut f\"{u}r Theoretische Physik und Astrophysik,
  Universit\"{a}t W\"{u}rzburg, Am Hubland,\\ D-97074 W\"{u}rzburg, Germany \and
  Max-Planck-Institut f\"{u}r Astrophysik,
  Karl-Schwarzschild-Str.\ 1,\\ D-85741 Garching, Germany}

\date{Received / Accepted}

\titlerunning{A localised subgrid scale model for fluid dynamical simulations I}
\authorrunning{W. Schmidt et al.}

\abstract{
	We present a one-equation subgrid scale model that evolves the turbulence energy
	corresponding to unresolved velocity fluctuations in large eddy simulations. 
	The model is derived in the context of the
	Germano consistent decomposition of the hydrodynamical equations.
	The eddy-viscosity closure for the rate of energy transfer from resolved toward
	subgrid scales is localised by means of a dynamical procedure for the
	computation of the closure parameter. Therefore,
	the subgrid scale model applies to arbitrary flow geometry and evolution.
	For the treatment of microscopic viscous dissipation a semi-statistical approach is used,
	and the gradient-diffusion hypothesis is adopted for turbulent transport.
	\emph{A priori} tests of the localised eddy-viscosity closure and the
	gradient-diffusion closure are made by analysing data from direct numerical
	simulations. As an \emph{a posteriori} testing case, the large eddy simulation
	of thermonuclear combustion in forced isotropic turbulence is discussed.
	We intend the formulation of the subgrid scale model in this paper
	as a basis for more advanced applications in numerical simulations of
	complex astrophysical phenomena involving turbulence.
}

\maketitle

\section{Introduction}

In the last decade, the significance of turbulence in
various astrophysical phenomena from stellar to
cosmological scales has been recognised.  In retrospect, this is
hardly surprising, since virtually all the matter in the Universe is
fluid, whereas the solid state is encountered as a rare
exception. Moreover, both the integral length $L$ and the
characteristic velocity $V$ in astrophysical systems are quite large
compared to terrestial standards, while the viscosity $\nu$ is
comparable to what is found for liquids or gas on the Earth. For this
reason, the dimensionless \emph{Reynolds number}
\begin{equation}
	\mathrm{Re}=LV/\nu
\end{equation}
becomes very large. A typical figure is $\mathrm{Re}\sim 10^{14}$ for turbulent 
stellar convection.

From the computational point of view, the number of degrees of freedom in a fluid dynamical system
 is given by the relation \citep[see][]{LanLifVI}
\begin{equation}
	\label{eq:landau}
	N\sim\left(L/\eta_{\mathrm{K}}\right)^{3}\sim\mathrm{Re}^{9/4}.
\end{equation}
The length scale $\eta_{\mathrm{K}}$ is called the \emph{Kolmogorov scale} and
specifies the smallest dynamically relevant length scale. Due to the
restriction of the CFL time step for compressible flow, the total number
of operations required to compute the evolution over one sound crossing 
time is of the order
\begin{equation}
	N_{\mathrm{cr}}\sim N^{4/3}\sim\mathrm{Re}^{3}.
\end{equation}
Hence, $N_{\mathrm{cr}}\sim 10^{42}$ operations would be required to
solve the problem of stellar convection in a direct numerical simulation.

The relevance of the Landau criterion~\ref{eq:landau} has been
questioned recently on the grounds of the \emph{intermittency} of
turbulence \citep{KritNor05}.  In fact, the number of degrees of
freedom $N$ refers to the ensemble of turbulent flow realisations.  At
any particular instant of time, however, turbulent dynamics is
concentrated in regions of fractal dimension $D<3$.  Topologically,
these regions can be either vortex filaments in subsonic flow or
shocklets in the case of supersonic turbulence.  The fractal dimension
of vortex filaments is $D=1$, whereas $D=2$ for shocklets.  According
to the $\beta$ model \citep[see][ Sect.~8.5]{Frisch}, the effective
number of degrees of freedom is then given by
\begin{equation}
	N_{\mathrm{eff}}\sim \mathrm{Re}^{3D/(D+1)}.
\end{equation}
If an adaptive numerical scheme with maximal efficiency in tracking
the intermittent turbulent regions were applied, the total
computational cost could be lowered substantially in comparison to a
direct numerical simulation on a static grid.  For example, it would
appear feasible to treat subsonic turbulence up to a Reynolds number
of $10^{10}$ with an anelastic adaptive code on high-end platforms in
the near future.  On account of the limited efficiency of adaptive
schemes, however, the actual constraint might be lower. Apart from
that, gravity, magnetohydrodynamic effects and, possibly, reaction
networks increase the work load dramatically in astrophysical
applications.  Moreover, we have $N_{\mathrm{cr}}\sim
N_{\mathrm{eff}}N^{1/3}\sim\mathrm{Re}^{11/4}$ for the particularly
interesting case of supersonic turbulence. As a consequence, even with
sophisticated adaptive schemes, it remains intractable to
resolve completely the turbulent fluid dynamics encountered in
astrophysics.

In large eddy simulations (LES), on the other hand, only a limited
number of degrees of freedom, which correspond to the largest scales
of the system, is treated explicitly.  For the turbulent dynamics on
smaller scales, a so-called \emph{subgrid scale model} is
utilised. Among astrophysicists, the most often used subgrid
scale (SGS) model is \emph{numerical dissipation}. This means that all
fluctuations on length scales smaller than the resolution $\Delta$ of
the numerical grid are smoothed out and it is assumed that the dynamics
on length scales larger than $\Delta$ are more or less independent of
the smaller scales. This point of view is motivated by the second
similarity hypothesis of \citet{Kolmog41}, which holds that the actual
mechanism of dissipation is insignificant, provided that there is
sufficient scale separation.  In other words, on length scales
$l\ga\Delta$, it is unimportant whether energy dissipation is caused
by the microscopic viscosity $\nu$ at the length scale
$\eta_{\mathrm{K}}$ or by numerical effects at the cutoff length
$\Delta$. The notion of numerical dissipation has been exhaustively
investigated for the piece-wise parabolic method (PPM) proposed by
\citet{ColWood84}, which is one of the most popular finite-volume
schemes applied in astrophysics. At least for statistically stationary
isotropic turbulence, the numerical dissipation of the PPM appears to
work well as an implicit SGS model \citep{SyPort00,SchmHille05a}.

For the treatment of transient or inhomogeneous turbulent flow,
however, an explicit SGS model becomes mandatory. 
One of the most prominent examples in contemporary
theoretical astrophysics are numerical simulations of turbulent
combustion in type Ia supernovae \citep{HilleNie00}.  In this case,
the velocity scale associated with SGS turbulence determines the
effective propagation speed of flame fronts \citep{NieHille95}. Part
II of this paper will be dedicated to the problem of type Ia supernova
simulations. The exchange of energy between resolved and
subgrid scales is expected to become dynamically significant in
the case of highly intermittent turbulence, for instance,
in collapsing turbulent gas clouds in the interstellar medium
\citep{Larson03,LowKless04}.

In this paper (paper I), we present a general framework
for the formulation of SGS models based upon the \emph{filtering
approach} of \citet{Germano92}.  The mathematical operation of
filtering smoothes the flow on length scales smaller than the
prescribed numerical resolution $\Delta$. Consequently, a scale
separation is introduced, where the smoothed density, velocity, etc.\
are identified with the resolved quantities computed in LES.  We have
extended this formalism to compressible flows, using the Reynolds
stress model proposed by \citet{Canuto97} as a guideline. Next we
discuss the one-equation SGS turbulence energy model
\citep{Schumann75,Sagaut}. For the energy transfer across the
numerical cutoff, we introduce a \emph{localised} eddy-viscosity
closure which makes use of the dynamical procedures introduced by
\citet{GerPio91}.  Hence, the SGS model becomes independent of \emph{a
priori} structural assumptions, in particular, whether the resolved
flow is homogeneous or stationary. However, it remains a necessary
condition that turbulent regions become locally isotropic on length
scales comparable to the numerical cutoff length $\Delta$, because the
linear eddy-viscosity closure presumes alignment between the
turbulence stress and the rate-of-strain tensors. In the future,
multi-parameter closures for the turbulent energy transfer which are
not subject to this restriction might be adapted. For the still more
complicated non-local transport of kinetic energy on subgrid scales,
we use a simple gradient-diffusion closure. In contrast to what has
been suggested in the literature \citep[][ Sect.~10.3]{Pope}, we find
that the optimal diffusivity parameter is larger by about one order of
magnitude than the SGS viscosity parameter, i.e. the turbulent kinetic
Prandtl number is large compared to unity.  Both the localised
eddy-viscosity and the statistical gradient-diffusion closure,
respectively, were tested by means of data from simulations of forced
compressible turbulence. As a case study, we present the LES of
turbulent combustion in a periodic box.  Although gravitational
effects on subgrid scales, in principle, can be incorporated into the
model as well, in paper I we restrict the detailed formulation and
application to the case of negligible gravity. However, a simple
closure which accounts for unresolved buoyancy effects in simulations
of thermonuclear supernovae will be discussed in paper II.

\section{Decomposition of the hydrodynamical equations}

Large eddy simulations pose the problem of scale separation. The
numerically computed flow can conceptually be defined by a set of smoothed
fields which correspond to the low-pass filtered physical flow realisation,
where $k_{\mathrm{c}}=\pi/\Delta$ is the cutoff wavenumber for a numerical
grid of resolution $\Delta$. A low-pass filter is a convolution operator
which is defined by
\begin{equation}
  \label{eq:q_flt}
  q(\vec{x},t) = \langle\ideal{q}\rangle_{G} \equiv
  \int\dd^{3}x'\,G(\vec{x}-\vec{x}',t)\ideal{q}(\vec{x}',t)
\end{equation}
for a particular kernel $G(\vec{x}-\vec{x}',t)$. The Fourier transform, the
so-called transfer function $\hat{G}(k,t)$, drops to zero for wavenumbers
$k\ga k_{\mathrm{c}}$. Consequently, only modes of wavenumbers less than
$k_{\mathrm{c}}$ contribute significantly to the filtered field $q(\vec{x},t)$.
The exact, unfiltered variable $\ideal{q}(\vec{x},t)$
corresponds to the limit $k_{\mathrm{c}}\rightarrow\infty$.
The filter operation smoothes out the fluctuations
$q'=\ideal{q}-q$ on length scales $l$ smaller than $\Delta$.
Albeit being mathematically determined by some dynamical equation,
$\ideal{q}(\vec{x},t)$ is generally not computable and therefore will be referred
to as an \emph{ideal} quantity.

The dynamical equation for the ideal velocity field is the generalisation of the Navier-Stokes
equation for compressible fluids \citep[see][]{LanLifVI}:
\begin{equation}
	\label{eq:vel_ideal}
	\frac{\ideal{\DD}}{\DD t}\ideal{\vec{v}} =
	\ideal{\vec{f}}.
\end{equation}
The differential operator on the left hand side is the \emph{Lagrangian} time derivate
\begin{equation}
	\frac{\ideal{\DD}}{\DD t} = \frac{\partial}{\partial t} + \ideal{\vec{v}}\cdot\vec{\nabla},
\end{equation}
and the effective force density acting upon the fluid is defined by
\begin{equation}
	\label{eq:force_eff}
	\ideal{\vec{F}} = \ideal{\rho}\ideal{\vec{f}} =
	-\vec{\nabla}\ideal{P} + \vec{\nabla}\cdot\ideal{\tens{\sigma}} +
	\ideal{\vec{F}}{^{\,(\mathrm{ext})}}.
\end{equation}
The first term on the right hand side is the pressure gradient, the second term
accounts for viscous dissipation and the third term is the total external force per unit volume
which encompasses gravitational and, possibly, stirring forces:
\begin{equation}
	\ideal{\vec{F}}{^{\,(\mathrm{ext})}} =
	\ideal{\rho}\ideal{\vec{g}} + \ideal{\vec{F}}{^{\,(\mathrm{s})}}.
\end{equation}
The \emph{viscous dissipation tensor} is defined by
\begin{equation}
	\label{eq:visc_diss_tens}
	\ideal{\sigma}_{ij} = 2\ideal{\rho}\ideal{\nu}\ideal{S}{_{ij}^{\ast}} =
	2\ideal{\rho}\ideal{\nu}\left(\ideal{S}_{ij} - \frac{1}{3}\ideal{d}\delta_{ij}\right),
\end{equation}
where $\nu$ is the microscopic viscosity of the fluid,
\begin{equation}
  \ideal{S}_{ij} = 
  \frac{1}{2}\left(\frac{\partial\ideal{v}_{i}}{\partial x_{j}} +
                   \frac{\partial\ideal{v}_{j}}{\partial x_{i}}\right),
\end{equation}
are the components of the \emph{rate-of-strain tensor} and
$\ideal{d}=\partial_{i}\ideal{v}_{i}$ is the divergence of the ideal flow.

Equation~(\ref{eq:vel_ideal}) can be written in a conservative form as
\begin{equation}
  \label{eq:vel_ideal_consrv}
  \frac{\partial}{\partial t}\ideal{\rho}\ideal{\vec{v}} + 
  \vec{\nabla}\cdot\ideal{\rho}\ideal{\vec{v}}\otimes\ideal{\vec{v}} =
  \ideal{\vec{F}}.
\end{equation}
The equality of the left hand side in both equations follows from
the continuity equation which expresses the conservation of mass
\begin{equation}
  \frac{\partial}{\partial t}\ideal{\rho} + 
  \vec{\nabla}\cdot\ideal{\rho}\ideal{\vec{v}} = 0  
\end{equation}

The goal of the filtering approach is the formulation of dynamical
equations for smoothed quantities which are amenable to the numerical computation.
We define the \emph{Favre} or mass-weighted filtered velocity field
\begin{equation}
\begin{split}
  \vec{v}(\vec{x},t) 
  &= \frac{\langle\ideal{\rho}\ideal{\vec{v}}\rangle_{G}}
       {\langle\ideal{\rho}\rangle_{G}} \\
  &= \frac{1}{\rho(\vec{x},t)}
  \int\dd^{3}x'\,G(|\vec{x}-\vec{x}'|)\ideal{\rho}(\vec{x}',t)
                                        \ideal{\vec{v}}(\vec{x}',t),
  \\
\end{split}
\end{equation}
where $\rho = \langle\ideal{\rho}\rangle_{G}$ is the filtered mass
density.
For a homogeneous and time-independent kernel,
the filter operation commutes with the Lagrangian derivative.
Favre filtering the dynamical equation~(\ref{eq:vel_ideal_consrv}), one obtains
\begin{equation}
  \frac{\partial}{\partial t}\rho\vec{v} + 
  \vec{\nabla}\cdot\langle\ideal{\rho}\ideal{\vec{v}}\otimes\ideal{\vec{v}}\rangle_{G} =
  \vec{F},
\end{equation}
where $\vec{F}=\langle\ideal{\vec{F}}\rangle_{G}$. The filtered equation
can be cast into a form analogous to equation~(\ref{eq:vel_ideal}),
\begin{equation}
	\label{eq:vel}
	\rho\frac{\DD}{\DD t}\vec{v} =
	\vec{F} +
	\vec{\nabla}\cdot\tau(\ideal{\rho}\ideal{\vec{v}},\ideal{\vec{v}}),
\end{equation}
by virtue of the \emph{generalised turbulence stress tensor}
\begin{equation}
	\tau(\ideal{\rho}\ideal{\vec{v}},\ideal{\vec{v}}) =
	-\langle\ideal{\rho}\ideal{\vec{v}}\otimes\ideal{\vec{v}}\rangle_{G} +
	\rho\vec{v}\otimes\vec{v}
\end{equation}
which \citet{Germano92} introduced without mass-weighing
for incompressible turbulence.
We will use $\tau_{ij}=\tau(\ideal{\rho}\ideal{v}_{i},\ideal{v}_{j})$ as a
shorthand notation for the components of the turbulence stress tensor.
The term $\partial_{j}\tau_{ij}$ in the momentum
equation
stems from the non-linear advection term in the Lagrangian time derivate
and can be interpreted as the stress exerted by
turbulent velocity fluctuations smoothed out by the filter. 
In the following, we will encounter a variety of $\tau$-terms. For this
reason, we define $\tau(\cdot,\cdot)$ as a generic bilinear form which maps any
pair of ideal fields to a mass-weighted smoothed field which is called
the \emph{generalised second moment}. The resulting field can be scalar,
vectorial or tensorial. Of course, the notion of a generalised moment applies to
moments of higher order as well.

A dynamical equation for the specific kinetic energy, $\ideal{k}=
\frac{1}{2}|\ideal{\vec{v}}|^{2}$, is readily obtained from the
contraction of equation~(\ref{eq:vel_ideal}) with the ideal
velocity field $\ideal{\vec{v}}$: 
\begin{equation}
	\label{eq:energy_kin_ideal}
	\ideal{\rho}\frac{\DD}{\DD t}\ideal{k} =
	\ideal{\vec{v}}\cdot\ideal{\vec{F}}.
\end{equation}
The mass-weighted filtered kinetic energy $k(\vec{x},t)$ is defined by
\begin{equation}
  	k(\vec{x},t) =
  	\frac{\langle\ideal{\rho}\ideal{\vec{k}}\rangle_{G}}
             {\langle\ideal{\rho}\rangle_{G}}.
\end{equation}
Filtering equation~(\ref{eq:energy_kin_ideal}) results in the
following equation for $k(\vec{x},t)$:
\begin{equation}
	\label{eq:energy_kin}
	\rho\frac{\DD}{\DD t}k - \vec{\nabla}\cdot\vec{\mathcal{F}}^{(\mathrm{kin})} =
	\vec{v}\cdot\vec{F} +
	\vec{\nabla}\cdot\left[\vec{v}\cdot\tau(\ideal{\rho}\ideal{\vec{v}},\ideal{\vec{v}})\right]
\end{equation}
In addition to the turbulence stress term on the right hand side, 
there is a non-local transport term which is given by the divergence
of the turbulent kinetic flux $\vec{\mathcal{F}}^{(\mathrm{kin})}$.
The flux is defined by the contraction
of the completely symmetric generalised third-order moment
$\tau(\ideal{\rho}\ideal{\vec{v}},\ideal{\vec{v}},\ideal{\vec{v}})$.
In component notation, we have
\begin{equation}
\begin{split}
  \label{eq:flux_kin}
  2\mathcal{F}_{i}^{(\mathrm{kin})}
  &= \tau_{ijj} \equiv
     \tau(\ideal{\rho}\ideal{v_{i}},\ideal{v_{j}},\ideal{v_{j}}) \\
  &= -\langle\ideal{\rho}\ideal{v_{i}}\ideal{v_{j}}\ideal{v_{j}}\rangle_{G} +
     \langle\ideal{\rho}\ideal{v_{i}}
     \ideal{v_{j}}\rangle_{G}v_{j} - 2\tau_{ij}v_{j} \\
\end{split}
\end{equation}

Since the filtered kinetic energy $k$ is a second-order moment of the
ideal velocity field, contributions from velocity fluctuations
on all scales are included. For this reason, $k$ differs from the
specific kinetic energy of the smoothed flow, $\frac{1}{2}|\vec{v}|^{2}$, and
\begin{equation}
	k_{\mathrm{turb}} = k - \frac{1}{2}|\vec{v}|^{2} = -\frac{1}{2}\tau_{ii}
\end{equation}
can be identified with the \emph{generalised turbulence energy} associated with scales
smaller than the characteristic length of the filter $\langle\ \rangle_{G}$.
The dynamical equation for $k_{\mathrm{turb}}$ is obtained by
subtracting
\begin{equation}
	\label{eq:energy_kin_res}
	\rho\frac{\DD}{\DD t}\left(\frac{1}{2}|\vec{v}|^{2}\right) =
	\vec{v}\cdot\left[\vec{F} +
	\vec{\nabla}\cdot\tau(\ideal{\rho}\ideal{\vec{v}},\ideal{\vec{v}})\right].
\end{equation}
from equation~(\ref{eq:energy_kin}). The result, in component notation, reads
\begin{equation}
	\label{eq:energy_turb_direct}
	\rho\frac{\DD}{\DD t}k_{\mathrm{turb}} - 
	\partial_{i}\mathcal{F}_{i}^{(\mathrm{kin})} =
	\tau(\ideal{\rho}\ideal{v}_{i},\ideal{v}_{j})S_{ij} -
	\tau(\ideal{v}_{i},\ideal{F}_{i}),
\end{equation}
where
$\tau(\ideal{v}_{i},\ideal{F}_{i})$ is the
contraction of the tensor
\begin{equation}
	\tau(\ideal{\vec{v}},\ideal{\vec{F}}) =
	-\langle\ideal{\vec{v}}\otimes\ideal{\vec{F}}\rangle_{G} +
	\vec{v}\otimes\vec{F}.
\end{equation}

In order to put equation~(\ref{eq:energy_turb_direct}) into a form which
is more adequate for modelling purposes, flux terms are split off on the
right hand side. Let us first substitute the definition of the 
effective force~(\ref{eq:force_eff}):
\begin{equation}
	\label{eq:energy_turb_source}
	\tau(\ideal{v}_{i},\ideal{F}_{i}) =
	-\tau(\ideal{v}_{i},\partial_{i}\ideal{P}) +
	\tau(\ideal{v}_{i},\partial_{j}\ideal{\sigma}_{ij}) +
	\tau(\ideal{v}_{i},\ideal{\vec{F}}{^{\,(\mathrm{ext})}}).
\end{equation}
The three resulting generalised moments respectively correspond to
pressure, viscous and external forces. Because stirring forces
usually supply energy on the integral length $L$ of the flow only, it follows that
$\tau(v_{i},F_{i}^{\,(\mathrm{s})})$ is negligibly small for
$\Delta\ll L$. The first and the second term on the right hand side of 
equation~(\ref{eq:energy_turb_source}) can be split as follows:
\begin{align}
	\label{eq:press_dilt_split}
	-\tau(\ideal{v}_{i},\partial_{i}\ideal{P}) &=
	-\partial_{i}\tau(\ideal{v}_{i},\ideal{P}) + \tau(\ideal{d},\ideal{P}), \\
	\label{eq:visc_diss_split}
	\tau(\ideal{v}_{i},\partial_{j}\ideal{\sigma}_{ij}) &=
	\partial_{j}\tau(\ideal{v}_{i},\ideal{\sigma}_{ij}) -
	\tau(\ideal{S}_{ij},\ideal{\sigma}_{ij}).
\end{align}
The new $\tau$-terms on the right hand side are, respectively, the
\emph{pressure-dilatation} and the \emph{rate of viscous dissipation}.
Substituting the expression for the viscous dissipation tensor~(\ref{eq:visc_diss_tens}),
it follows that
\begin{equation}
	\label{eq:visc_diss_full}
	\tau(\ideal{S}_{ij},\ideal{\sigma}_{ij}) =
	-\langle\ideal{\rho}\ideal{\nu}|\ideal{S}{^{\ast}}|^{2}\rangle_{G} -
	2\langle\ideal{\rho}\ideal{\nu}\ideal{S}{_{ij}^{\ast}}\rangle_{G}S_{ij}.
\end{equation}
The norm $|\ideal{S}{^{\ast}}|$ is defined by the total contraction
of the trace-free part of the rate of strain tensor:
\begin{equation}
	|\ideal{S}{^{\ast}}|^{2} = 
	2\ideal{S}{_{ij}^{\ast}}\ideal{S}{_{ij}^{\ast}} =
	|\ideal{S}|^{2}-\frac{2}{3}\ideal{d}{^{2}}.
\end{equation}
The norm of the rate-of-strain tensor 
is a probe of the velocity fluctuations on the smallest dynamical
length scales which are of the order of the Kolmogorov scale
$\eta_{\mathrm{K}}$. We shall assume that the characteristic
length of the filter is much greater than the Kolmogorov scale. In this case,
the rate of strain of the filtered velocity field is much less than
the rate of strain of the ideal velocity field, i.e.
$|\ideal{S}{^{\ast}}|^{2}\la|S^{\ast}|^{2}$. Consequently,
the first term on the right hand side of equation~(\ref{eq:visc_diss_full})
dominates the second term, and we can set
\begin{equation}
	\label{eq:visc_diss}
	\tau(\ideal{S}_{ij},\ideal{\sigma}_{ij}) \simeq
	-\langle\nu\ideal{\rho}|\ideal{S}{^{\ast}}|^{2}\rangle_{G}.
\end{equation}

Summarising, equation~(\ref{eq:energy_turb_direct}) can be written in the form
\begin{equation}
	\label{eq:energy_turb}
	\rho\frac{\DD}{\DD t}k_{\mathrm{turb}} - \mathfrak{D} =
	\Gamma + \Sigma - \rho(\lambda+\epsilon),
\end{equation}
where the source contributions on the right hand side are
\begin{align}
	\label{eq:turb_gravity}
	\Gamma &= \langle\ideal{\rho},\ideal{v}_{i}\ideal{g}_{i}\rangle_{G} - \rho v_{i}g_{i}, \\
	\Sigma &= \tau_{ij}S_{ij}, \\
	\rho\lambda &= -\langle\ideal{d}\ideal{P}\rangle_{G} + d P, \\
	\label{eq:turb_diss}
	\rho\epsilon &= \langle\ideal{\nu}\ideal{\rho}|\ideal{S}{^{\ast}}|^{2}\rangle_{G},
\end{align}
and the transport term $\mathfrak{D}$ is given by
\begin{equation}
	\mathfrak{D} =
	\frac{\partial}{\partial x_{i}}
	\left[\frac{1}{2}\tau_{ijj} + \mu_{i} +
	      \langle\ideal{\sigma}_{ij}\ideal{v}_{j}\rangle_{G}\right].
\end{equation}
The generalised moment
\begin{equation}
	\label{eq:flux_press}
	\vec{\mu} = -\langle\ideal{\vec{v}}\ideal{P}\rangle_{G}+\vec{v}P.
\end{equation}
accounts for the transport of turbulence energy due to pressure fluctuations.

For the internal energy, the filtered dynamical equation is
\begin{equation}
\begin{split}
	\label{eq:energy_int}
	\rho\frac{\DD}{\DD t}e_{\mathrm{int}}
	-&\vec{\nabla}\cdot\left[\vec{\mathcal{F}}^{(\mathrm{cond})} +  
          \tau(\ideal{\rho}\ideal{\vec{v}},\ideal{e}_{\mathrm{int}})\right] \\
	&= \rho\chi\vec{\nabla}T +
	Q - \langle\ideal{P}\ideal{d}\rangle_{G} - \rho\epsilon. \\
\end{split}
\end{equation}
The source term $Q$ accounts for heat generation by chemical or nuclear
reactions. 
The transport of heat due to turbulent temperature
fluctuations gives rise to the \emph{generalised conductive flux},
\begin{equation}
	\vec{\mathcal{F}}^{(\mathrm{cond})} = -\tau(\ideal{\rho}\ideal{\chi},\vec{\nabla}\ideal{T}) =
	\langle\ideal{\rho}\ideal{\chi}\ideal{T}\rangle_{G} -
	\rho\chi\vec{\nabla}T,
\end{equation}
for fluid of thermal conductivity $\ideal{\chi}$. The additional
transport term on the left hand of equation~(\ref{eq:energy_int}) side
arises from the transport of heat by turbulent advection. In the case
of buoyancy-driven turbulence, this transport mechanism is known as
convection.  The \emph{generalised convective flux} is defined by
\begin{equation}
	\vec{\mathcal{F}}^{(\mathrm{conv})} =
	\tau(\ideal{\rho}\ideal{\vec{v}},\ideal{h}) =
	\tau(\ideal{\rho}\ideal{\vec{v}},\ideal{e}_{\mathrm{int}}) + \vec{\mu},
\end{equation}
where $h=e_{\mathrm{int}}+P/\rho$ is the specific enthalpy.

Adding the budgets of the specific kinetic energy
$\frac{1}{2}|\vec{v}|^{2}$ and internal energy $e_{\mathrm{int}}$, we
obtain the total energy per unit mass on length scales $l\ga\Delta$,
i.e. $e_{\mathrm{tot}}=e_{\mathrm{int}}+\frac{1}{2}|\vec{v}|^{2}$. The
dynamical equation governing the evolution of $e_{\mathrm{tot}}$ is
\begin{equation}
\begin{split}
	\label{eq:energy_tot}
	\rho\frac{\DD}{\DD t}e_{\mathrm{tot}} 
	-& \vec{\nabla}\cdot\left[\vec{\mathcal{F}}^{(\mathrm{cond})} +
	\vec{\mathcal{F}}^{(\mathrm{conv})} - vP\right] \\
	&= \rho\chi\vec{\nabla}T + Q + \rho(\lambda + \epsilon) \\
	&\phantom{=}+ \vec{v}\cdot\left[\vec{g}+\vec{f}^{(\mathrm{s})} +
	\vec{\nabla}\cdot\tau(\ideal{\rho}\ideal{\vec{v}},\ideal{\vec{v}})\right]. \\
\end{split}
\end{equation}
This conservation law in combination with
equation~(\ref{eq:energy_turb}) extends the \emph{Germano consistent
decomposition} to compressible turbulence. The sum of internal energy
and kinetic energy on all scales is $e_{\mathrm{tot}} +
k_{\mathrm{turb}} = e_{\mathrm{int}} + k$.  Comparing
equations~(\ref{eq:energy_turb}) and (\ref{eq:energy_tot}), one can
see that $\rho(\lambda + \epsilon)$ accounts for the dissipation of
turbulence energy into internal energy by compression effects and
viscous dissipation, respectively.  The turbulence production term
$\Sigma$ is related to the energy transfer through the turbulence
cascade across the characteristic length of the filter. The
injection of energy due to buoyancy and the action of stirring forces
on length scales larger than $\Delta$ is given by
$\vec{v}\cdot[\vec{g}+\vec{f}^{(\mathrm{s})}]$, whereas the
interaction of gravitational potential energy fluctuations and
turbulence on length scales smaller than $\Delta$ is accounted for by
the term $\Gamma$.

From the discussion in this Section, it should become
clear that the presumed scale separation in LES is essentially based upon
the disentanglement of a variety of dynamical effects. This task is considerably
complicated by the non-linear interactions across the cutoff scale,
which become manifest in the various generalised moments
occurring in the equations~(\ref{eq:energy_kin}), (\ref{eq:energy_turb}),
(\ref{eq:energy_int}) and~(\ref{eq:energy_tot}).
Hence, one faces the problem of finding \emph{closures} for the generalised
moments in the decomposed dynamical equations.

In the simplest of all cases, a closure is a sufficiently convincing argument
for neglecting a certain term. This kind of closure is applied in many cases. 
In the proper sense, a closure is a more or less
tentative approximation which is made on grounds of heuristic physical
arguments. Two major categories of closures can be distinguished:
An algebraic closure is some function of filtered quantities only.
Usually, algebraic closures contain at least one free parameter. Depending on whether this
parameter is a constant or varies in space and time, the closure
is either statistical or localised.
On the other hand, dynamical closures determine generalised moments 
from additional dynamical equations. However,
these equations, in turn, entail closures for higher-order
generalised moments. This is the problem of the infinite hierarchy of
equations for filtered quantities. Inevitably, the hierarchy must
be truncated at some point with the help of algebraic closures.

\section{The subgrid scale turbulence energy model}

In LES, numerical solutions for the filtered quantities $\rho$, $P$, $T$, $\vec{v}$ and
$e_{tot}$ have to be computed from the continuity equation for $\rho$, the momentum equation~(\ref{eq:vel}),
the energy conservation law~(\ref{eq:energy_tot}) and the equation of state. 
The filter operation naturally introduces a cutoff which is related to the
numerical resolution. Since the filtering in LES is not necessarily explicit
but sometimes inherent to the numerical scheme, we will subsequently
use the generic notation $\langle\ \rangle_{\mathrm{eff}}$. For example,
\begin{equation}
  \label{eq:vel_eff_def}
  \vec{v}(\vec{x},t) =
  \frac{\langle\ideal{\rho}\ideal{\vec{v}}(\vec{x},t)\rangle_{\mathrm{eff}}}
       {\langle\ideal{\rho}\rangle_{\mathrm{eff}}},
\end{equation}
is the mass-weighted velocity field which is to be computed numerically.
For finite precision, the numerical solution actually corresponds
to a whole ensemble of exact flow realisations. In this regard,
one can think of a reduction of the number of degrees of freedom
due to filtering.

The length scales smaller than the characteristic length $\Delta_{\mathrm{eff}}$ 
of the effective filter are the \emph{subgrid scales}
(SGS). The scales $l\ga\Delta_{\mathrm{eff}}$, on the other hand, are
numerically resolved. A complete set of closures for the generalised
moments in the dynamical equations constitutes the \emph{subgrid scale model}.
A general SGS model which includes dynamical equations for the moments of
second and third order was formulated by \citet{Canuto94}. Unfortunately, the computational cost 
of solving the whole set of dynamical equations for the generalised moments is considerable.
Moreover, the problem of stability appears to be non-trivial.

The SGS model which will be discussed in the following involves the
solution of the dynamical
equation for the subgrid scale turbulence energy only.
The definition of the density of SGS turbulence energy is as follows:
\begin{equation}
	K_{\mathrm{sgs}} \equiv \frac{1}{2}\rho q_{\mathrm{sgs}}^{2} =
	-\frac{1}{2}\tau_{ii} =
	\frac{1}{2}\left[\langle\ideal{\rho}|\ideal{\vec{v}}|^{2}\rangle_{\mathrm{eff}} -
	                 \rho|\vec{v}|^{2}\right].
\end{equation}
The magnitude of SGS velocity fluctuations is given by $q_{\mathrm{sgs}}$.
The equation governing the evolution of the specific turbulence energy
$k_{\mathrm{sgs}}=K_{\mathrm{sgs}}/\rho$ is just
equation~(\ref{eq:energy_turb}) with the filter 
$\langle\ \rangle_{\mathrm{eff}}$. For the various second-order moments
in this equation we will invoke standard algebraic closures. Hence,
the implementation of the SGS model requires the solution of only
one additional dynamical equation. The inherent limitations of the
simple closures are in part compensated by localisation. Thereby, the
SGS model becomes basically independent of \emph{a priori} model
parameters which presume certain flow properties such as homogeneity.
In essence, the SGS model which will be formulated is robust and
particularly well suited for complex flow geometries and transients, although requiring
relatively little computational resources.

There are two different sources of SGS turbulence production.
The first one is the SGS gravity term $\Gamma_{\mathrm{sgs}}$~(\ref{eq:turb_gravity}) which accounts
for the conversion of potential into kinetic energy and vice versa
due to correlations between SGS fluctuations of the velocity and gravity.
A putative closure for SGS buoyancy in reactive flows
will be presented in paper II. The other production term is
the rate of energy transfer $\Sigma_{\mathrm{sgs}}=\tau_{ij}S_{ij}$
across the length scale $\Delta_{\mathrm{eff}}$ due to non-linear
turbulent interactions. In general, energy transfer is the primary
source of SGS turbulence. A common closure is based on the 
\emph{eddy-viscosity hypothesis} for the the trace-free 
part of the SGS turbulence stress tensor \citep[cf.][ Sect.~10.1.]{Pope}:
\begin{equation}
  \label{eq:turb_visc_cl}
  \tau_{ij}^{\ast} \circeq
  2\rho\nu_{\mathrm{sgs}}S_{ij}^{\ast} =
  2\rho\nu_{\mathrm{sgs}}\left(S_{ij} - \frac{1}{3}d\delta_{ij}\right),
\end{equation}
where
\begin{equation}
  \label{eq:turb_stress_cl}
  \tau_{ij}^{\ast} =
  \tau_{ij} - \frac{1}{3}\tau_{ii}\delta_{ij} =
  \tau_{ij} + \frac{2}{3}K_{\mathrm{sgs}}\delta_{ij}.
\end{equation}
This closure is formulated analogously to the viscous stress tensor in
a Newtonian fluid. The eddy viscosity $\nu_{\mathrm{sgs}}$ is
assumed to be proportional to the product of the characteristic length
$\Delta_{\mathrm{eff}}$ of the numerical scheme and the characteristic
velocity of SGS turbulence (cf.\ \citealp[ Sect.~4.3]{Sagaut}, and
\citealp[ Sect.~13.6.3]{Pope}), i.\ e.,
\begin{equation}
  \label{eq:visc_sgs}
  \nu_{\mathrm{sgs}} \circeq
  C_{\nu}\Delta_{\mathrm{eff}}k_{\mathrm{sgs}}^{1/2} =
  \ell_{\nu}q_{\mathrm{sgs}}.
\end{equation}
The length scale $\ell_{\nu}=C_{\nu}\Delta_{\mathrm{eff}}/\sqrt{2}$
is thus associated with SGS turbulence production.

Among the dissipation terms, the rate of viscous dissipation
$\epsilon_{\mathrm{sgs}}$ defined in equation~(\ref{eq:turb_diss})
dominates in subsonic turbulent flows.  Assuming a Kolmogorov
spectrum, the mean SGS turbulence energy corresponding to a sharp
spectral cut-off can be related to the mean rate of dissipation:
\begin{equation}
  \langle k_{\mathrm{sgs}}\rangle =
  \int_{\pi/\Delta}\overset{\infty}{E}(k)\dd k =
  \frac{3}{2}C\langle\epsilon_{\mathrm{sgs}}\rangle^{2/3}
  \left(\frac{\pi}{\Delta}\right)^{-2/3}.
  \end{equation}
Hence, assuming that $C\approx 1.65$ \citep{YeuZhou97},
\begin{equation}
  \langle\epsilon_{\mathrm{sgs}}\rangle =
  \Sigma\left(\frac{3C}{2}\right)^{-3/2}
  \frac{\langle k_{\mathrm{sgs}}\rangle^{3/2}}{\Delta} \approx
  0.81\frac{\langle k_{\mathrm{sgs}}\rangle^{3/2}}{\Delta}.
\end{equation}
Conjecturing that the above relation
also holds locally
\citep[cf.][ Sect.~13.6.3]{Pope}, we have
\begin{equation}
  \label{eq:diss_close}
  \epsilon_{\mathrm{sgs}} \circeq
  C_{\epsilon}
  \frac{k_{\mathrm{sgs}}^{3/2}}{\Delta_{\mathrm{eff}}} =
  \frac{q_{\mathrm{sgs}}^{3}}{\ell_{\epsilon}},
\end{equation}
where $\ell_{\epsilon}=2\sqrt{2}\Delta_{\mathrm{eff}}/C_{\epsilon}$
and $C_{\epsilon}\sim 1$. Basically, equation~(\ref{eq:diss_close})
implies that SGS eddies of kinetic energy $\sim
q_{\mathrm{sgs}}^{2}$ are dissipated on a time scale $\sim
\ell_{\epsilon}/q_{\mathrm{sgs}}$.

Pressure dilatation poses severe difficulties because one needs to
find the correlations between pressure fluctuations and compression or
rarefaction of the fluid. The first-order hypothesis is that kinetic
energy is dissipated if the fluid is contracting $(d < 0)$. In the
opposite case ($d > 0$), internal energy is converted into mechanical
energy which produces turbulence.  This line of reasoning leads to the
closure proposed by \citet{Dear73}:
\begin{equation}
  \lambda_{\mathrm{sgs}} \circeq C_{\lambda}k_{\mathrm{sgs}}d.
\end{equation}
Unfortunately, numerical tests reveal that this closure is extremely
crude. Since compressibility effects tend to diminish toward smaller
length scales, the above closure will do for the LES of subsonic
turbulence.  In the case of supersonic turbulence, however,
$\lambda_{\mathrm{sgs}}$ becomes more significant. Alternative
closures for pressure-dilatation are described in \citet{Canuto97}.

A customary algebraic closure for the transport term in
equation~(\ref{eq:energy_turb}) is the \emph{gradient-diffusion
hypothesis} \citep[cf.][ Sect.~4.3]{Sagaut}\footnote{ Also known as
\emph{Kolmogorov-Prandtl relation}. }
\begin{equation}
  \label{eq:diff_close}
  \mathfrak{D}_{\mathrm{sgs}} \circeq
  \frac{\partial}{\partial x_{k}}\rho
  C_{\kappa}\Delta_{\mathrm{eff}}k_{\mathrm{sgs}}^{1/2}
  \frac{\partial k_{\mathrm{sgs}}}{\partial x_{k}} =
  \frac{\partial}{\partial x_{k}}\rho
  \ell_{\kappa}q_{\mathrm{sgs}}^{2}
  \frac{\partial q_{\mathrm{sgs}}}{\partial x_{k}}.
\end{equation}
The characteristic length scale of diffusion is defined by
$\ell_{\kappa}=C_{\kappa}\Delta_{\mathrm{eff}}/\sqrt{2}$ and the SGS
diffusivity is given by $\kappa_{\mathrm{sgs}} =
\ell_{\kappa}q_{\mathrm{sgs}}$. The notion of a turbulent diffusivity
of kinetic energy stems from the analogy to the thermal diffusion of
internal energy on microscopic scales. This analogy also suggests the
definition of a \emph{ kinetic Prandtl number},
\begin{equation}
  \mathrm{Pr}_{\,\mathrm{kin}} = \frac{\nu_{\mathrm{sgs}}}{\kappa_{\mathrm{sgs}}} =
  \frac{C_{\nu}}{C_{\kappa}}.
\end{equation}

Summarising, if gravitational effects on subgrid scales
are negligible, we obtain the following dynamical equation for the
SGS turbulence energy:
\begin{equation}
\begin{split}
  \label{eq:energy_sgs}
  \frac{\DD}{\DD t}k_{\mathrm{sgs}}
  -&\frac{1}{\rho}\vec{\nabla}\cdot\left(\rho C_{\kappa}\Delta_{\mathrm{eff}}
    k_{\mathrm{sgs}}^{1/2}\vec{\nabla}k_{\mathrm{sgs}}\right) \\
  &= C_{\nu}\Delta_{\mathrm{eff}}k_{\mathrm{sgs}}^{1/2}|S^{\ast}|^{2} -
     \left(\frac{2}{3}+C_{\lambda}\right)k_{\mathrm{sgs}}d -
     C_{\epsilon}\frac{k_{\mathrm{sgs}}^{3/2}}{\Delta_{\mathrm{eff}}}.
\end{split}
\end{equation}
The remaining problem is the determination of the closure parameters
$C_{\kappa}$, $C_{\nu}$, $C_{\lambda}$ and $C_{\epsilon}$ which are
dimensionless similarity parameters, i.e. the values become asymptotically
scale invariant in statistically stationary isotropic turbulence.

\section{Closure parameters}

In this section, methods for the calculation of closure parameters will
be discussed. In particular, we will present a so-called
\emph{dynamical procedure} for the computation of the eddy-viscosity
parameter $C_{\nu}$. Originally introduced by engineers in order to
improve the performance of simple SGS models such as the Smagorinsky
model for turbulent flows near walls, the application of dynamical
procedures for the localised computation of closure parameters has
turned out to be a powerful tool for the treatment of inhomogeneous
and non-stationary turbulence. For this reason, we adapted a procedure
proposed by \citet{KimMen99} for the LES of turbulent combustor
flows. Using data from simulations of forced isotropic turbulence, we
found that this procedure yields a significantly better match with the
rate of production than the statistical closure with a constant
parameter. For the parameter of dissipation, $C_{\epsilon}$, we
propose a semi-statistical solution: A time-dependent value is
determined from the energy budget of the resolved flow in extended
spatial regions. Regarding the non-local transport, the
gradient-diffusion closure produces satisfactory results if the
parameter $C_{\kappa}$ is determined appropriately.

\subsection{Production}
\label{sc:prod}

The rate of production $\Sigma_{\mathrm{sgs}}$ corresponds to
dissipation of kinetic energy on resolved scales due to the effect of
subgrid scale turbulence. Pictorially, unresolved eddies drain energy
from larger eddies at the rate $\Sigma_{\mathrm{sgs}}$. This idea
motivated the \emph{eddy-viscosity} closure~(\ref{eq:turb_visc_cl})
for $\Sigma_{\mathrm{sgs}}$. Extending further the analogy between viscous
and turbulent dissipation, an experimental assertion known as the
\emph{law of finite dissipation} could be carried over to the
production of SGS turbulence: If, in a large eddy simulation of
turbulent flow, all the control parameters are kept the same except
for $\Delta_{\mathrm{eff}}$, which is
lowered as much as possible, the energy dissipation per unit mass,
$\Sigma_{\mathrm{sgs}}$, behaves in a way consistent with a finite
positive limit\footnote{ The formulation is the same as in
\citet{Frisch}, beginning of Sect. 5.  }. This suggests that the
parameter $C_{\nu}$ in the definition of the turbulent
viscosity~(\ref{eq:visc_sgs}) becomes asymptotically scale-invariant
in the limit $\Delta_{\mathrm{eff}}/L\rightarrow 0$ and assumes a
universal value in the stationary limit.

We verified this hypothesis by analysing data from numerical
simulations of forced isotropic turbulence. The driving force which
supplies energy on the characteristic length scale $L$ is modelled by
a stochastic process in Fourier space \citep{EswaPope88,Schmidt04}.
Under the action of this force, the flow evolves on the characteristic
time $T$ which is called the \emph{large-eddy time scale}.  In the
statistically stationary limit, the flow velocity is of the order
$V=L/T$. In addition, the weight of solenoidal (divergence-free)
relative to dilatational (rotation-free) components of the force field
can be varied by setting the control parameter $\zeta$ in the range
between $1$ and $0$. Choosing different characteristic Mach numbers
$V/c_{0}$, where $c_{0}$ is the initial sound speed, and values of
$\zeta$, we performed several simulations using the piece-wise
parabolic method (PPM) with $N=432^{3}$ grid cells \citep{ColWood84}.
A realistic equation of state for electron-degenerate matter was used
in these simulations \citep[see][]{Rein01} and the numerical dissipation of PPM
provided an implicit subgrid scale model.

It is possible to evaluate generalised moments from the simulation data
on a length scale which is large compared to the cutoff scale $\Delta$.
To that end, let us introduce new smoothed fields $\rho^{<}$ and
$\vec{v}^{<}$ which are associated with a basis filter $\langle\
\rangle_{<}$ of characteristic length $\Delta_{<}$:
\begin{alignat}{2}
	\rho^{<} &= \langle\ideal{\rho}\rangle_{<}, \qquad
	\rho^{<}\vec{v}^{<} &= \langle\ideal{\rho}\ideal{\vec{v}}\rangle_{<}.
\end{alignat}
If $\Delta_{<}$ is sufficiently large compared to
$\Delta_{\mathrm{eff}}$, then $q\equiv \langle\ideal{q}\rangle_{<}
\simeq \langle q\rangle_{<}$ for any quantity $q$.  This property of
low-pass filters becomes immediately apparent in spectral space in
which the filter operation is just a multiplication of Fourier modes
with the transfer function. Consequently, the turbulence stress tensor
at the level of the basis filter is approximately given by
\begin{equation}
	\label{eq:stress_basis}
	\tau_{<}(\ideal{\rho}\ideal{\vec{v}},\ideal{\vec{v}}) \approx
	\tau_{<}(\rho\vec{v},\vec{v}) =
	-\langle\rho\vec{v}\otimes\vec{v}\rangle_{<} +
	\rho^{<}\vec{v}^{<}\otimes\vec{v}^{<}.
\end{equation}

\begin{table}
  \caption{Closure parameters for selected flow realisations from three different
    simulations of forced isotropic turbulence}
  \label{tb:closr_parmtrs}
  \begin{center}
    \begin{tabular}{c l c r c c c r}
      \hline
      $V/c_{0}$ & $\zeta$ & $t/T$ & $\Delta_{<}/\Delta$ & $\langle C_{\nu}\rangle$ &
      $C_{\kappa}^{(\mathrm{vec})}$ & $C_{\kappa}^{(\mathrm{scl})}$ &
      $\mathrm{Pr}_{\,\mathrm{kin}}$ \\
      \hline\hline
      0.42 & 1.0  & 2.5 &  6.9 & 0.0512 & 0.0653 & 0.401 &  7.8 \\
      0.66 & 0.75 & 2.0 &  9.6 & 0.0447 & 0.0611 & 0.401 &  9.0 \\
      0.66 & 0.75 & 4.0 &  6.8 & 0.0476 & 0.0649 & 0.422 &  8.7 \\
      0.66 & 0.75 & 9.0 &  6.5 & 0.0451 & 0.0661 & 0.529 & 11.7 \\
      1.39 & 0.2  & 3.5 & 16.1 & 0.0370 & 0.0986 & 0.481 & 13.0 \\
      1.39 & 0.2  & 6.0 &  6.4 & 0.0422 & 0.0806 & 0.515 & 12.2 \\
      \hline
    \end{tabular}
  \end{center}
\end{table}

Evaluating $\tau_{<}(\rho\vec{v},\vec{v})$, it is possible to determine the
eddy-viscosity parameter from the rate of energy transfer across
the length scale $\Delta_{<}$ :
\begin{equation}
	C_{\nu}^{<} =
	\frac{\Sigma_{<}}
	     {\rho^{<}\Delta_{<}\sqrt{k_{\mathrm{turb}}^{<}}\,|S^{<\,\ast}|},
\end{equation}
where $\Sigma_{<} = \tau_{<}^{\ast}(\rho v_{i},v_{j})S_{ij}^{<}$ and
$\rho^{<}k_{\mathrm{turb}}^{<}\simeq
-\frac{1}{2}\tau_{<}(\rho v_{i},v_{i})$. We computed $C_{\nu}^{<}$
from several flow realisations.  A sample of average values for the
whole domain is listed in Table~\ref{tb:closr_parmtrs}. We found
$\langle C_{\nu}^{<}\rangle\approx 0.05$ for developed turbulence, in
agreement with the literature \citep{Pope}. Only in the case of
transonic flow with a mostly compressive driving force the
eddy-viscosity parameters appears to be systematically
lower. Fig.~\ref{fg:prod3d} (a) shows a visualisation of the
turbulence energy isosurfaces given by
$k_{\mathrm{turb}}^{<}=0.25V^{2}$ with the contour sections of the
dimensionless rate of energy transfer
$\tilde{\Sigma}_{<}=(T/\rho_{0}V^{2})\Sigma_{<}$ for $V/c_{0}=0.66$ at
time $t=4T$. The corresponding contours obtained with the
eddy-viscosity closure and $C_{\nu}^{<}=0.0476$ are plotted in panel
(b).  Clearly, the rate of energy transfer is not well
reproduced. Although there is a significant correlation of about
$0.8$, the magnitude of spatial variations is greatly reduced.

\begin{figure*}[thb]
  \begin{center}
    \vspace{80mm}
    \mbox{\subfigure[Explicit]{\texttt{\hspace{30mm}fig\_transf.png\hspace{30mm}}}\qquad
          \subfigure[Statistical closure]{\texttt{\hspace{30mm}fig\_closr\_ave.png\hspace{30mm}}}}\\
    \vspace{80mm}
    \mbox{\subfigure[Localised closure]{\texttt{\hspace{30mm}fig\_closr\_locl.png\hspace{30mm}}}\qquad
          \subfigure[Localised closure excluding
          backscattering]{\texttt{\hspace{30mm}fig\_closr\_locl\_nb.png\hspace{30mm}}}}
    \caption{Isosurfaces of the turbulence energy $k_{\mathrm{turb}}^{<}=0.25V^{2}$ with contour sections
    	of the dimensionless rate of energy transfer. The flow realisation is
    	taken from a simulation with characteristic Mach number $V/c_{0}=0.66$ at time $t=4T$.}
    \label{fg:prod3d}
  \end{center}
\end{figure*}

In fact, $C_{\nu}^{<}$ exhibits spatiotemporal variations comparable
to the mean value. In consequence, the assumption of a constant
eddy-viscosity closure parameter is not valid. However, the
information about the variation of $C_{\nu}$ is  not
available in a LES. A solution to this problem can be found by means
of a similarity hypothesis which relates the energy transfer across
different scales. Let us consider a length scale
$\Delta_{\mathrm{T}}$ which is somewhat larger than
$\Delta_{<}$.  Introducing a suitably defined filter
operation $\langle\ \rangle_{\mathrm{T}}$ of characteristic length
$\Delta_{\mathrm{T}}$, the turbulence stress
$\tau_{\mathrm{T}}(\rho\vec{v},\vec{v})$ is given by an expression
analogous to the right hand side of equation~(\ref{eq:stress_basis}):
\begin{equation}
	\tau_{\mathrm{T}}(\rho^{<}\vec{v}^{<},\vec{v}^{<}) =
	-\langle\rho^{<}\vec{v}^{<}\otimes\vec{v}^{<}\rangle_{\mathrm{T}} +
	\rho^{(\mathrm{T})}\vec{v}^{(\mathrm{T})}\otimes\vec{v}^{(\mathrm{T})}
\end{equation}
Here $\rho^{(\mathrm{T})}=\langle\rho^{<}\rangle_{\mathrm{T}}$ and
$\vec{v}^{(\mathrm{T})}=\langle\rho^{<}\vec{v}^{<}\rangle_{\mathrm{T}}/
\langle\rho^{<}\rangle_{\mathrm{T}}$.
The stress tensors associated with the length scales
$\Delta_{\mathrm{T}}$ and $\Delta_{<}$, respectively, are
related by an identity which Germano \citep{Germano92} originally
formulated for incompressible turbulence:
\begin{equation}
	\label{eq:germano_id}
	\tau_{\mathrm{T}}(\rho\vec{v},\vec{v})=
	\langle\tau_{<}(\rho\vec{v},\vec{v})\rangle_{\mathrm{T}} +
	\tau_{\mathrm{T}}(\rho^{<}\vec{v}^{<},\vec{v}^{<}).
\end{equation}
The first term on the right-hand side is the filtered turbulence
stress tensor associated with the length scale $\Delta_{<}$, whereas
the second term accounts for the turbulence stress on intermediate
length scales in between $\Delta_{<}$ and $\Delta_{\mathrm{T}}$. For
small scaling ratios
$\gamma_{\mathrm{T}}=\Delta_{\mathrm{T}}/\Delta_{<}$, there is
significant correlation not only between
$\tau_{\mathrm{T}}(\rho\vec{v},\vec{v})$ and
$\langle\tau_{<}(\rho\vec{v},\vec{v})\rangle_{\mathrm{T}}$, but also
between $\tau_{\mathrm{T}}(\rho\vec{v},\vec{v})$ and
$\tau_{\mathrm{T}}(\rho^{<}\vec{v}^{<},\vec{v}^{<})$. In particular,
this was demonstrated by \citet{LiuMen94} from velocity measurements
in round jets.

Based upon these experimental findings, \citet{KimMen99} proposed a
similarity hypothesis for the eddy-viscosity parameter:
\begin{equation}
	C_{\nu}^{<} = C_{\nu}^{(\mathrm{T})} =
	\frac{\Sigma^{(\mathrm{T})}}
	     {\rho^{(\mathrm{T})}\Delta_{\mathrm{T}}k_{\mathrm{T}}^{1/2}|S^{\ast\,(\mathrm{T})}|},
\end{equation}
where $|S^{\ast\,(\mathrm{T})}|$ is the norm of
\begin{equation}
	S_{ij}^{(\mathrm{T})} =
	\frac{1}{2}\left[\partial_{j}v_{i}^{(\mathrm{T})}+\partial_{i}v_{j}^{(\mathrm{T})}\right],
\end{equation}
and $\Sigma^{(\mathrm{T})} = \tau_{\mathrm{T}}^{\ast}(\rho^{<}v_{i}^{<},v_{j}^{<})S_{ij}^{(\mathrm{T})}$.
The specific turbulence energy $k_{\mathrm{T}}$ corresponding to intermediate
velocity fluctuations in between the basis and the test filter level
is defined by
\begin{equation}
  \label{eq:energy_test}
	\rho^{(\mathrm{T})}k_{\mathrm{T}} =
	-\frac{1}{2}\tau_{\mathrm{T}}(\rho^{<}v_{i}^{<},v_{i}^{<}) =
	-\frac{1}{2}\tau_{\mathrm{T}}(\rho v_{i},v_{i}) - 
	\langle\rho^{<}k_{\mathrm{turb}}^{<}\rangle_{\mathrm{T}}.
\end{equation}
The second expression for $k_{\mathrm{T}}$ follows from the
contraction of the Germano identity~(\ref{eq:germano_id}).  Thus, the
parameter $C_{\nu}^{<}$ for the eddy-viscosity closure at the level of
the basis filter is determined by probing the flow at the length scale
$\Delta_{\mathrm{T}}>\Delta_{<}$.  This is why $\langle\
\rangle_{\mathrm{T}}$ is called a \emph{test filter}.

\begin{figure*}[thb]
  \begin{center}
    \mbox{\subfigure{\includegraphics[width=8.5cm]{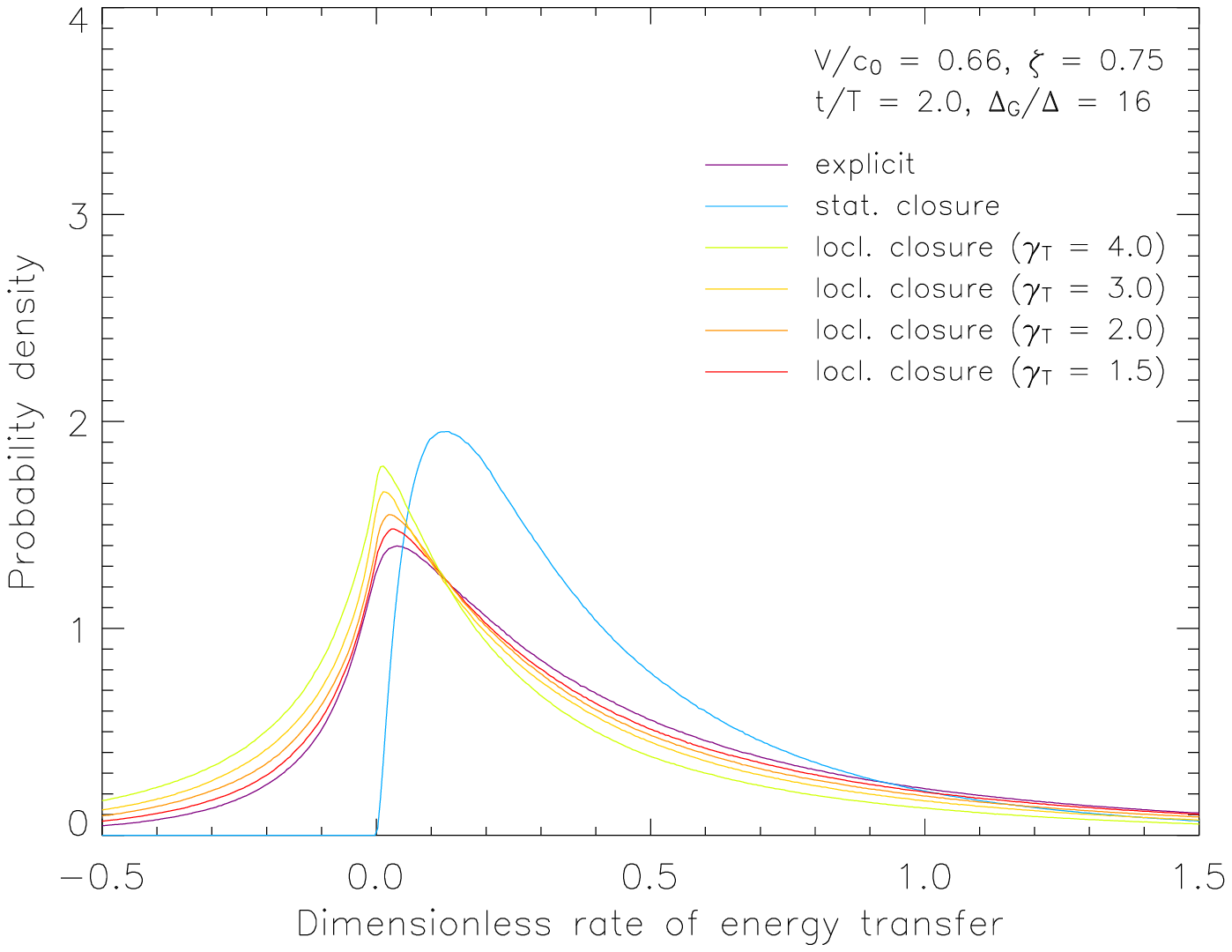}}
          \subfigure{\includegraphics[width=8.5cm]{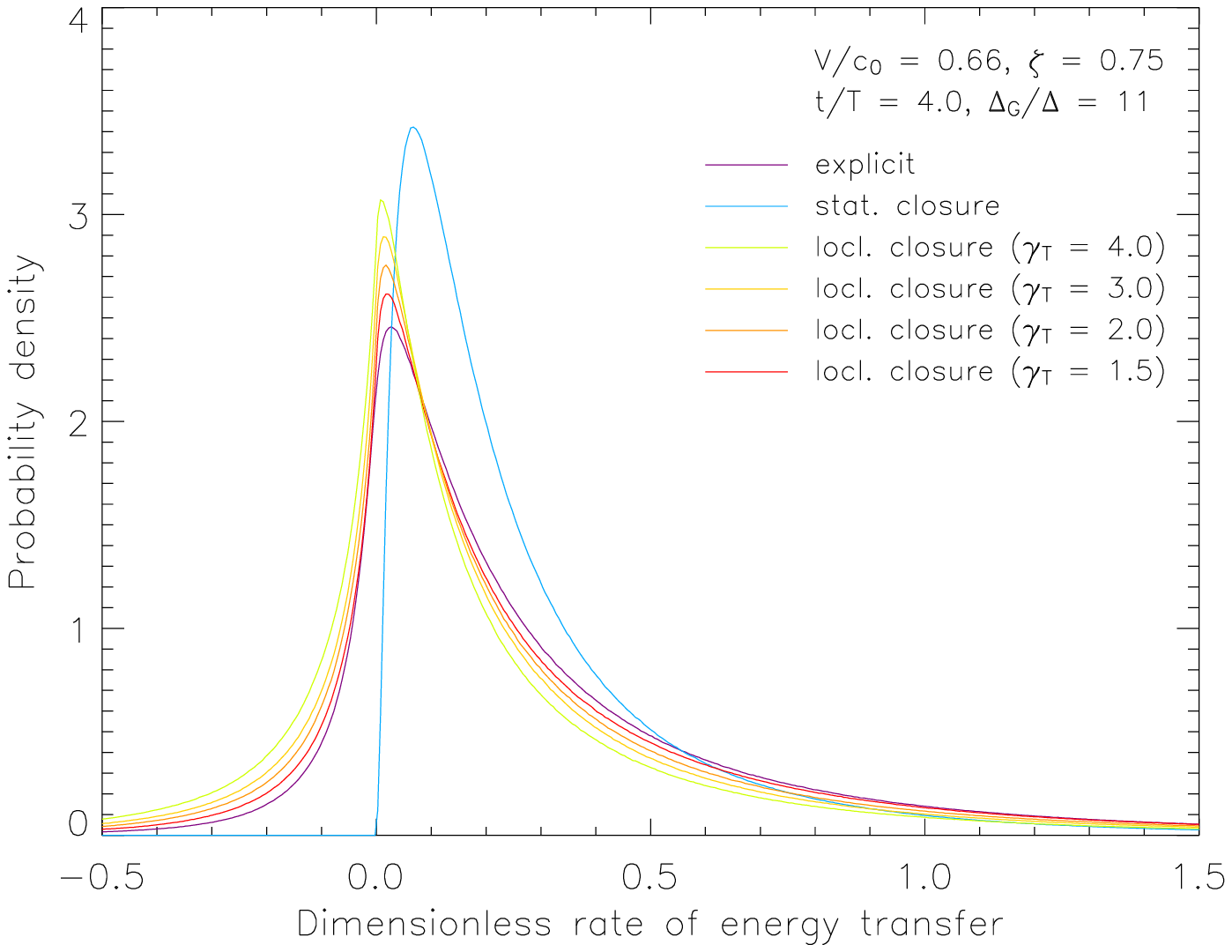}}}
    \caption{Probability density functions for the rate of energy
             transfer $\tilde{\Sigma}_{<}$ across the length scale $\Delta_{<}$
             in dimensionless scaling at two different instants of time.}
    \label{fg:transf_pdf}
  \end{center}
\end{figure*}

Using data from the simulations of forced isotropic
turbulence, we tested the proposition made above by computing
explicitly the rate of energy transfer across a certain length scale
$\Delta_{<}$ and comparing it to the eddy-viscosity closure with the
closure parameter calculated at test filter levels for different
scaling ratios $\gamma_{\mathrm{T}}$.  In order to apply
approximation~(\ref{eq:stress_basis}), we had to choose a basis filter
length $\Delta_{<}$ which was at least an order of magnitude larger
than the resolution $\Delta$ in the simulations. On the other hand, a
sufficient range of inertial length scales greater than $\Delta_{<}$
is required for the test filter operation. These requirements
substantially constrained the choice of $\Delta_{<}$.  Further
complications come from the so-called bottleneck effect which causes a
distortion of the energy spectrum function for wave numbers close to
the cutoff at $k_{\mathrm{c}}=\pi/\Delta$
\citep{DobHau03,HauBrand04}. A detailed discussion of the
kinetic energy spectrum functions and, particularly, the bottleneck
effect in turbulence simulations with PPM is given in
\citet{SchmHille05a}. As one can see in Fig.~\ref{fg:transf_pdf},
the match between the probability density functions of the
dimensionless rate of energy transfer and the corresponding localised
closure is substantially better than for the closure with constant
eddy-viscosity parameter. This is highlighted by the statistical
moments listed in Table~\ref{tb:closr_stat_momt}. In particular, the
variance of the energy transfer is largely underestimated by the
statistical closure. This also becomes apparent from the
three-dimensional visualisations in Fig.~\ref{fg:prod3d} (b) and (c),
respectively, which suggest that variations of the energy transfer are
flattened by a wide margin in the case of a constant eddy-viscosity
parameter, while the localised closure reproduces local extrema quite
well.  On the other hand, it appears that the characteristic length
$\Delta_{\mathrm{T}}$ of the test filter should not be chosen too
large in relation to $\Delta^{<}$. Otherwise the mean of the energy
transfer will be systematically underestimated
(Fig.~\ref{fg:transf_pdf} and Table~\ref{tb:closr_stat_momt}).

\begin{table}[htb]
  \caption{Statistical moments of the dimensionless rate of
    energy transfer for the probability density functions
    plotted in Fig.~\ref{fg:transf_pdf} (b).}
  \label{tb:closr_stat_momt}
  \begin{center}
    \begin{tabular}{l c c c}
      \hline
      computation & $\langle\tilde{\Sigma}_{<}\rangle$ & 
      $\sigma(\tilde{\Sigma}_{<})$ & $\mathrm{skew}(\Sigma_{<})$ \\
      \hline\hline
      explicit                                  & 0.330 & 0.526 & 3.70 \\
      stat.~closure                             & 0.293 & 0.327 & 3.53 \\
      locl.~closure ($\gamma_{\mathrm{T}}=1.5$) & 0.313 & 0.552 & 3.71 \\
      locl.~closure ($\gamma_{\mathrm{T}}=2.0$) & 0.273 & 0.540 & 3.88 \\
      locl.~closure ($\gamma_{\mathrm{T}}=3.0$) & 0.234 & 0.525 & 3.73 \\
      locl.~closure ($\gamma_{\mathrm{T}}=4.0$) & 0.183 & 0.504 & 3.51 \\
      \hline
    \end{tabular}
  \end{center}
\end{table}

\begin{figure}[thb]
  \begin{center}
    \resizebox{\hsize}{!}{\includegraphics[width=17cm]{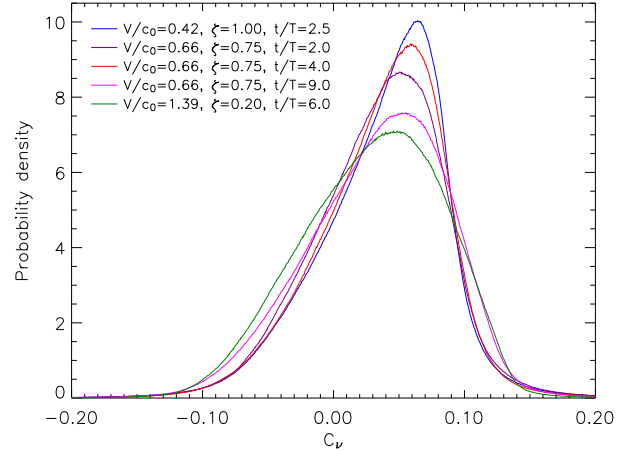}}
    \caption{Probability density functions for the localised eddy viscosity
      parameter calculated from different flow realisations.}
    \label{fg:c_prod}
  \end{center}
\end{figure}

The variability of $C_{\nu}^{(\mathrm{T})}$ is illustrated by
the probability density functions plotted in Fig.~\ref{fg:c_prod}.
The similarity of the functions suggest a fairly robust behaviour of
$C_{\nu}^{(\mathrm{T})}$ for driven isotropic turbulence.  In a
fraction of roughly $15$ to $20\,\%$ of the domain, negative values of
the closure parameter are found which are commonly interpreted as
inverse energy transfer from length scales smaller than
$\Delta_{\mathrm{T}}$ toward larger scales. This phenomenon, which is
also know as backscattering, is predicted by turbulence
theory. However, as we shall argue in Sect.~\ref{sc:burn_box},
backscattering introduces numerical difficulties in combination with
PPM. But panel (d) in Fig.~\ref{fg:prod3d} demonstrates that the
localised closure is superior even when negative values of the
eddy-viscosity are suppressed.

For the application in LES, the basis filter corresponds to the
effective filter introduced in the previous Sect., and the
test filter is applied to the computed fields $\rho(\vec{x},t)$ and
$\vec{v}(\vec{x},t)$. Then we have 
\begin{equation}
	\label{eq:c_prod_locl}
	C_{\nu} =
	\frac{\tau_{\mathrm{T}}^{\ast}(\rho v_{i},v_{j})S_{ij}^{(\mathrm{T})}}
	     {\Delta_{\mathrm{T}}k_{\mathrm{T}}^{1/2}|S^{\ast\,(\mathrm{T})}|}.
\end{equation}
The characteristic length scale of SGS turbulence production,
$\ell_{\nu}=$, depends on the scaling ratio
$\gamma_{\mathrm{T}}=\Delta_{\mathrm{T}}/\Delta_{\mathrm{eff}}$ and,
consequently, $\ell_{\nu}C_{\nu}\Delta_{\mathrm{eff}}/\sqrt{2}$ is
proportional to the parameter $\beta=\Delta_{\mathrm{eff}}/\Delta$.
\citet{SchmHille05a} determined $\beta\approx 1.6$ for the
statistically stationary turbulent regime in simulations with PPM.

\subsection{Dissipation}

The localised closure for the rate of production works because the
energy transfer across a certain cutoff wavenumber is mostly
determined by interactions between Fourier modes within a narrow band
around the cutoff.  Concerning the rate of dissipation
$\epsilon_{\mathrm{sgs}}$, we encounter an entirely different problem.
In fact, viscous dissipation takes place on length scales which are of
the order of the Kolmogorov scale
$\eta_{\mathrm{K}}\la\Delta_{\mathrm{eff}}$.  There is no obvious
similarity between the dissipation on resolved scales (due to SGS
turbulence) and the dissipation on subgrid scales (due to microscopic
viscosity). The simplest of all SGS models, which is known as the
Smagorinsky model, assumes a local equilibrium between the dissipation
on resolved and subgrid scales, respectively.  However, it is the very
point of the SGS turbulence energy model that such a balance does not
hold locally. Nevertheless, the mean rate of energy transfer
can be related to the rate of viscous dissipation in the case of
homogeneous turbulence. If the flow is inhomogeneous, equilibrium
might be assumed at least for some nearly homogeneous regions. Thus,
we attempt to determine the closure parameter $C_{\epsilon}$ from the
averaged energy budget on the test filter level for a suitably chosen
flow region.

The method is loosely based on the variational
approach of \cite{GhoLund95}. They subtracted the test-filtered SGS
turbulence energy equation~(\ref{eq:energy_sgs}) from the
corresponding equation for the turbulence energy at the level
of the test filter in order to determine $C_{\epsilon}$ as a
function of both space and time. Our approach is an intermediate
one, where spatially averages energy equations are considered.
For the mean SGS turbulence energy, averaging equation~(\ref{eq:energy_sgs})
yields
\begin{equation}
  \label{eq:energy_sgs_ave}
  \left\langle\rho\frac{\DD}{\DD t}k_{\mathrm{sgs}}\right\rangle = 
  \langle\tau_{ik}S_{ik}\rangle - 
  \left\langle\rho(\lambda_{\mathrm{sgs}}+\epsilon_{\mathrm{sgs}})\right\rangle.
\end{equation}
Here it is assumed that the surface contributions from the transport
term cancel out or at least can be neglected. Since
\begin{equation}
  \left\langle\rho\frac{\DD}{\DD t}k_{\mathrm{sgs}}\right\rangle =
  \left\langle\frac{\partial}{\partial t}\rho k_{\mathrm{sgs}}\right\rangle +
  \underbrace{\left\langle\frac{\partial}{\partial x_{i}}
              \rho v_{i}k_{\mathrm{sgs}}\right\rangle}_{\simeq 0} \simeq
  \frac{\dd}{\dd t}\langle K_{\mathrm{sgs}}\rangle,
\end{equation}
we also neglect the effect of advection by the resolved flow.
The turbulence energy density associated with the characteristic scale of the
test filter is defined by the trace of the Germano identity~(\ref{eq:germano_id})
\begin{equation}
  \label{eq:energy_sgs_test_ave}
  -\frac{1}{2}\tau_{\mathrm{T}}(\ideal{\rho}\overset{\infty}{v_{i}},\overset{\infty}{v_{i}}) =
  -\frac{1}{2}\langle\tau_{ii}\rangle_{\mathrm{T}} + \frac{1}{2}\tau_{\mathrm{T}}(\rho v_{i},v_{i}) =
  \langle K_{\mathrm{sgs}}\rangle_{\mathrm{T}} + K_{\mathrm{T}},
\end{equation}
where
$K_{\mathrm{T}}=\rho^{(\mathrm{T})}k_{\mathrm{T}}$~(\ref{eq:energy_test}).
The spatial average of the turbulence
energy~(\ref{eq:energy_sgs_test_ave}) is given by the following
dynamical equation:
\begin{equation}
\begin{split}
  \frac{\partial}{\partial t}\langle K_{\mathrm{sgs}}+K_{\mathrm{T}}\rangle =& 
  \left\langle\tau_{\mathrm{T}}(\ideal{\rho}\overset{\infty}{v_{i}},\overset{\infty}{v_{k}})
     S_{ik}^{(\mathrm{T})}\right\rangle \\
  &-\left\langle\rho(\lambda_{\mathrm{sgs}}+\epsilon_{\mathrm{sgs}})+
     \rho^{(\mathrm{T})}(\lambda_{\mathrm{T}}+\epsilon_{\,\mathrm{T}})\right\rangle. \\
\end{split}
\end{equation}

Equations~(\ref{eq:energy_sgs_ave}) and~(\ref{eq:energy_sgs_test_ave}) in combination with the Germano
identity~(\ref{eq:germano_id}) imply
\begin{equation}
\begin{split}
  \label{eq:energy_test_ave}
  \frac{\dd}{\dd t}\langle K_{\mathrm{T}}\rangle =& 
  \left\langle\tau_{\mathrm{T}}(\rho v_{i},v_{k})S_{ik}^{(\mathrm{T})} +
     \langle\tau_{ik}\rangle_{\mathrm{T}}S_{ik}^{(\mathrm{T})} -
     \tau_{ik}S_{ik}\right\rangle \\
  &-\left\langle\rho^{(\mathrm{T})}(\lambda_{\mathrm{T}}+\epsilon_{\,\mathrm{T}})\right\rangle.
\end{split}
\end{equation}
Substituting the eddy-viscosity closures for the various
production terms on the right-hand side, the above equation becomes
\begin{equation}
  \begin{split}	
     \frac{\dd}{\dd t}\langle K_{\mathrm{T}}\rangle \simeq
     & \underbrace{\left\langle\rho^{(\mathrm{T})} C_{\nu}\Delta_{\mathrm{T}}k_{\mathrm{T}}^{1/2}
         |S^{\ast\,(\mathrm{T})}|^{2}\right\rangle -
         \frac{2}{3}\left\langle K_{\mathrm{T}}
         d^{(\mathrm{T})}\right\rangle}_{\mathrm{(I)}}\\
     & - \langle\rho^{(\mathrm{T})}\lambda_{\mathrm{T}}\rangle
       + \langle\rho^{(\mathrm{T})}\epsilon_{\,\mathrm{T}}\rangle\\
     & + \underbrace{\left\langle\langle\rho\nu_{\mathrm{sgs}}S_{ik}^{\ast}\rangle_{\mathrm{T}}
         S_{ik}^{\ast\,(\mathrm{T})} -
         \rho\nu_{\mathrm{sgs}}|S^{\ast}|^{2}\right\rangle}_{\mathrm{(II)}}\\
     & - \frac{2}{3}\underbrace{\left\langle\langle
         K_{\mathrm{sgs}}\rangle_{\mathrm{T}}d^{(\mathrm{T})}-
         K_{\mathrm{sgs}}d\right\rangle}_{\mathrm{(III)}}.\\
  \end{split}                 
\end{equation}
Due to the large number of filtered quantities, the complete numerical
computation of the source terms in the above equation would be
rather demanding.  For this reason, we drop the contributions (II) and (III) 
while only retaining (I), which is presumably the most significant contribution
to the rate of energy transfer across $\Delta_{\mathrm{T}}$.
Then $\langle K_{\mathrm{T}}\rangle$ is approximately given by
\begin{equation}
\begin{split}
  \label{eq:energy_test_ave_approx}
  \frac{\dd}{\dd t}\langle K_{\mathrm{T}}\rangle =&
  \left\langle\rho^{(\mathrm{T})} C_{\nu}\Delta_{\mathrm{T}}k_{\mathrm{T}}^{1/2}
    |S^{\ast\,(\mathrm{T})}|^{2}\right\rangle \\
  &-\frac{2}{3}\left\langle\rho^{(\mathrm{T})}(k_{\mathrm{T}}d^{(\mathrm{T})}+
                         \lambda_{\mathrm{T}})\right\rangle -
  \langle\rho^{(\mathrm{T})}\epsilon_{\,\mathrm{T}}\rangle.\\
\end{split}
\end{equation}

Invoking the closure dimensional closure~(\ref{eq:diss_close}) both for the
SGS rate of dissipation $\epsilon_{\mathrm{sgs}}$ and the
rate of dissipation at the test filter level, we obtain the
following expression for the mean rate of dissipation on length
scales in between $\Delta_{\mathrm{eff}}$ and $\Delta_{\mathrm{T}}$:
\begin{equation}
  \label{eq:diss_test1}
  \langle\rho^{(\mathrm{T})}\epsilon_{\,\mathrm{T}}\rangle \circeq
  \frac{C_{\mathrm{\epsilon}}}{\Delta_{\mathrm{T}}}\left\langle\rho^{(\mathrm{T})}
    \left(\frac{\langle\rho k_{\mathrm{sgs}}\rangle_{\mathrm{T}}}{\rho^{(\mathrm{T})}} +
          k_{\mathrm{T}}\right)^{3/2} -
    \gamma_{\mathrm{T}}\rho k_{\mathrm{sgs}}^{3/2}\right\rangle
\end{equation}
Furthermore, setting 
\begin{equation}
  \label{eq:pd_test}
  \lambda_{\mathrm{T}}\circeq
  C_{\mathrm{\lambda}}k_{\mathrm{T}}d^{(\mathrm{T})},
\end{equation}
the closure parameter $C_{\epsilon}$ is determined by
\begin{equation}
\begin{split}
  \label{eq:c_diss_ave}
  C_{\epsilon} = 
  & -\left[\frac{\dd}{\dd t}\langle K_{\mathrm{T}}\rangle -
  \left\langle C_{\nu}\rho^{(\mathrm{T})}\Delta_{\mathrm{T}}k_{\mathrm{T}}^{1/2}
          |S^{\ast\,(\mathrm{T})}|^{2}\right\rangle +
  \left(\frac{1}{3}+C_{\lambda}\right)\left\langle K_{\mathrm{T}}
                                                  d^{(\mathrm{T})}\right\rangle\right]\\
  & \times\Delta_{\mathrm{T}}\left\langle\rho^{(\mathrm{T})}
  \left(\frac{\langle\rho k_{\mathrm{sgs}}\rangle_{\mathrm{T}}}{\rho^{(\mathrm{T})}} +
        k_{\mathrm{T}}\right)^{3/2} -
  \gamma_{\mathrm{T}}\rho k_{\mathrm{sgs}}^{3/2}\right\rangle^{-1}.\\
\end{split}
\end{equation}
Contrary to the eddy-viscosity parameter $C_{\nu}$ which varies both in
space and time $C_{\epsilon}$ 
is a time-dependent constant for a
suitably chosen spatial region. For homogeneous turbulence, there is only
one region encompassing the whole domain of the flow.  In a
stratified medium, it is appropriate to average horizontally. 
Then $C_{\epsilon}$ varies with depth. For turbulent
combustion problems, such as type Ia supernova explosions, one can
distinguish fuel, the burning zone and the burned
material within. For each of these three regions a value of the
dissipation parameter is calculated as a function of time. 
For the pressure-dilatation parameter $C_{\lambda}$,
on the other hand, we preliminarily assume the constant,
time-independent value $C_{\lambda}=-\frac{1}{5}$ for subsonic
turbulence \citep[see][]{FurTab97}.

\subsection{Diffusion}

As in the case of the energy transfer, we shall first consider the problem of non-local
transport at the level of a basis filter of characteristic length $\Delta_{<}$ which
is large compared to the numerical cutoff length. The generalised kinetic flux~(\ref{eq:flux_kin})
is given by
\begin{equation}
\begin{split}
  \mathcal{F}_{i}^{(kin)\,<} =&
    -\frac{1}{2}\langle\rho v_{i}v_{j}v_{j}\rangle_{<} +
     \frac{1}{2}v_{i}^{<}\langle\rho v_{j}v_{j}\rangle_{<} \\
  &+\langle\rho v_{i}v_{j}\rangle_{<}v_{j}^{<} -
    \rho^{<}v_{i}^{<}v_{j}^{<}v_{j}^{<}\\
\end{split}
\end{equation}
and the pressure diffusion flux~\ref{eq:flux_press} reads
\begin{equation}
	\vec{\mu} = -\langle P\vec{v}\rangle_{<}+P^{<}\vec{v}^{<}.
\end{equation}
Assuming that the total flux vector $\vec{\mathcal{F}}^{(kin)\,<} + \vec{\mu}^{<}$
is aligned with the turbulence energy gradient $\vec{\nabla}k_{\mathrm{turb}}^{<}$,
the gradient-diffusion closure can be written as follows:
\begin{equation}
	\label{eq:grad_diff_close}
	\vec{\mathcal{F}}^{(kin)\,<} + \vec{\mu}^{<} \circeq
	C_{\kappa}^{<}\Delta_{<}\sqrt{k_{\mathrm{turb}}^{<}}\,\vec{\nabla}k_{\mathrm{turb}}^{<}.
\end{equation}
Contracting the above relation with $\vec{\nabla}k_{\mathrm{turb}}^{<}$ and averaging
over the domain of the flow, one obtains
\begin{equation}
	\label{eq:c_diff_vec}
	C_{\kappa}^{(\mathrm{vec})\,<} =
	\frac{\langle\vec{\mathcal{F}}^{(\mathrm{kin})\,<} + \vec{\mu}^{<}\rangle\cdot\vec{\nabla}k_{\mathrm{turb}}^{<}}
	     {\Delta_{<}\left\langle\sqrt{k_{\mathrm{turb}}^{<}}\,
                                             |\vec{\nabla}k_{\mathrm{turb}}^{<}|^{2}\right\rangle}.
\end{equation}

\begin{figure}[htb]
  \begin{center}
    \resizebox{\hsize}{!}{\subfigure{\includegraphics[width=17cm]{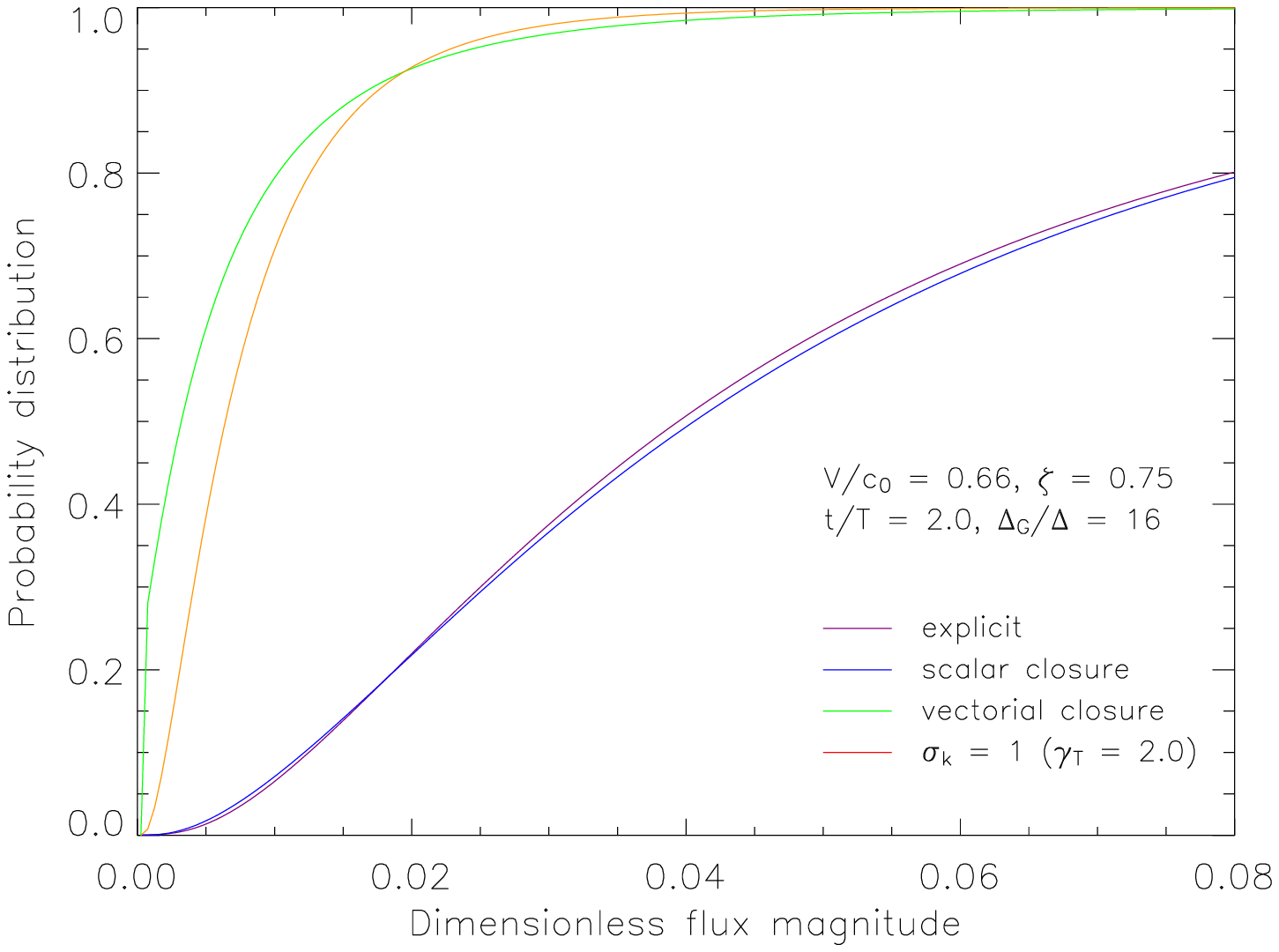}}}
    \resizebox{\hsize}{!}{\subfigure{\includegraphics[width=17cm]{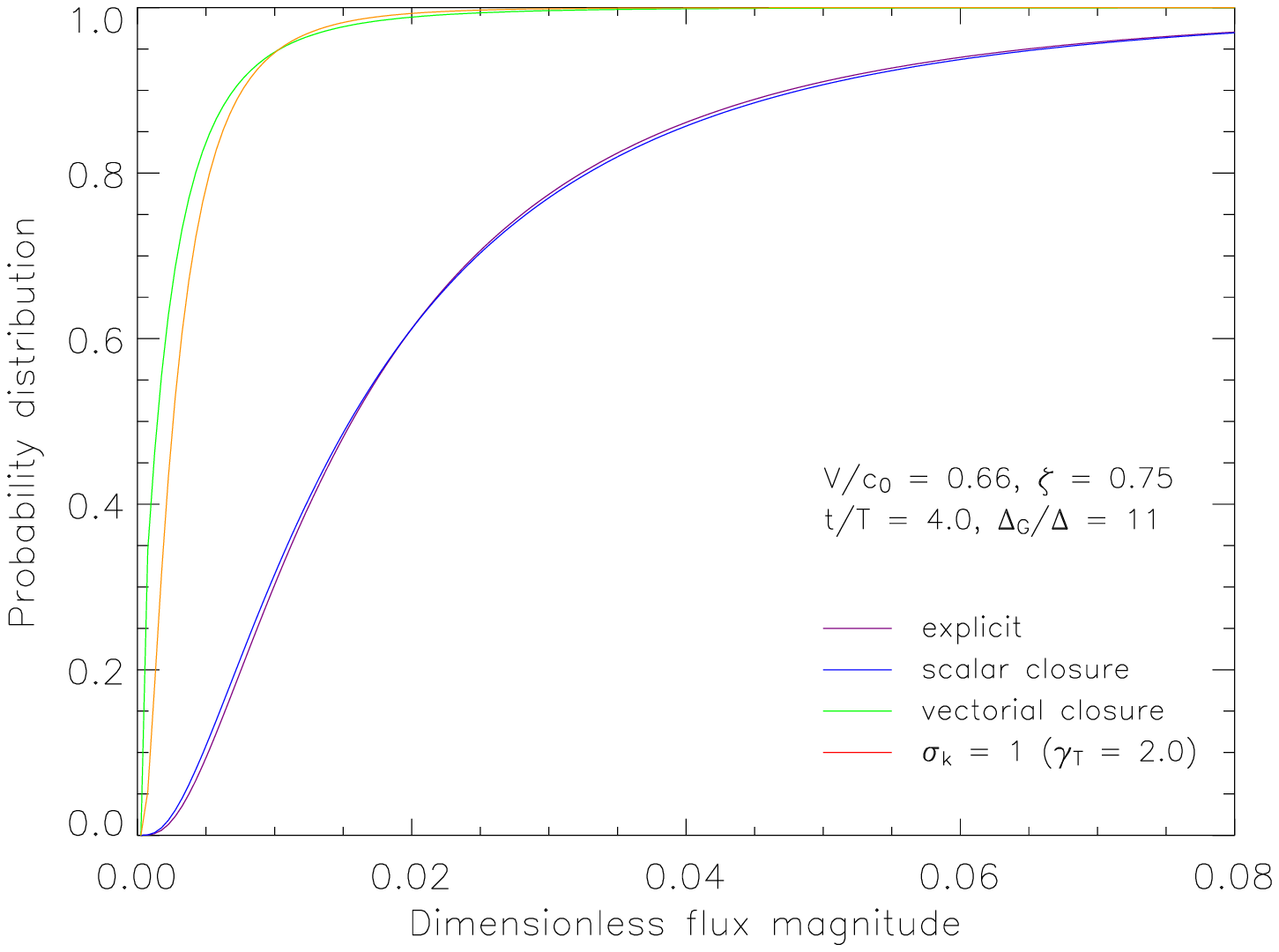}}}
    \caption{Probability distribution functions for $|\vec{\mathcal{F}}^{(kin)\,<} + \vec{\mu}^{<}|$
    	and the corresponding gradient-diffusion closures with different
    	turbulent diffusivity parameters. }
    \label{fg:diff}
  \end{center}
\end{figure}

A sample of values for $C_{\kappa}^{(\mathrm{vec})<}$ is listed in
Table~\ref{tb:closr_parmtrs}.  In agreement with a turbulent kinetic
Prandtl number of the order unity, $C_{\kappa}^{(\mathrm{vec})<}$ is
of the same order of magnitude as the closure parameter $C_{\nu}^{<}$
\citep[see][ Sect.~10.3]{Pope}.  Contour sections of the flux magnitude
$|\vec{\mathcal{F}}^{(kin)\,<}+\vec{\mu}^{<}|$ and the corresponding
closure at the $k_{\mathrm{turb}}^{<}=0.25$ isosurfaces for $V/c_{0}=0.66$ at time
$t=4T$ are shown in Fig.~\ref{fg:diff3d}. However, as one can see
from a comparison of the panels (a) and (b), the closure
underestimates the diffusive flux by about an order of magnitude.
Even more clearly, this is demonstrated by the probability
distribution functions plotted in Fig.~\ref{fg:diff}. We also
investigated the hypothesis of setting the turbulent diffusivity
parameter equal to the localised eddy-viscosity parameter
\citep[see][ Sect.~4.3]{Sagaut,KimMen99}.  Since negative diffusivity
would induce numerical instability, we truncated the diffusivity parameter at zero,
i.e. $C_{\kappa}^{<}=C_{\nu}^{(\mathrm{T})\,+}$.  The resulting
visualisation in panel (c) of Fig.~\ref{fg:diff3d} and the
corresponding graph in Fig.~\ref{fg:diff}, however, show very little
if any improvement compared to the statistical closure.

\begin{figure*}[thb]
  \begin{center}
    \vspace{79mm}
    \mbox{\subfigure[Explicit]{\texttt{\hspace{30mm}fig\_diff.png\hspace{30mm}}}\qquad
          \subfigure[$C_{\kappa}=0.065$ (vectorial)]{\texttt{\hspace{30mm}fig\_diff\_closr\_vec.png\hspace{30mm}} }}\\
    \vspace{79mm}
    \mbox{\subfigure[$C_{\kappa}=C_{\nu}^{(\mathrm{T})\,+}$]{\texttt{\hspace{30mm}fig\_diff\_locl\_nb.png\hspace{30mm}}}\qquad
	  \subfigure[$C_{\kappa}=0.422$ (scalar)]{\texttt{\hspace{30mm}fig\_diff\_closr\_scl.png\hspace{30mm}}} }
    \caption{Turbulence energy isosurfaces as in Fig.~\ref{fg:prod3d} with contour sections
    	of the dimensionless flux magnitude of turbulent transport.}
    \label{fg:diff3d}
  \end{center}
\end{figure*}

The reason for the discrepancies is the flawed assumption of alignment
between the turbulent flux vector and the energy gradient. Setting
\begin{equation}
	\label{eq:c_diff_scl}
	C_{\kappa}^{(\mathrm{scl})<} =
	\frac{\langle|\vec{\mathcal{F}}^{(\mathrm{kin})\,<} + \vec{\mu}^{<}|\rangle}
	     {\Delta_{<}\left\langle|\vec{\nabla}k_{\mathrm{turb}}^{<}|
                                             \sqrt{k_{\mathrm{turb}}^{<}}\right\rangle},
\end{equation}
where an equality of the flux magnitude but not the direction is
presumed, results in significantly larger turbulent diffusivity (see
Table~\ref{tb:closr_parmtrs}).  In particular, panel (d) in
Fig.~\ref{fg:diff3d} and the probability distribution functions
shown in Fig.~\ref{fg:diff} reveal a very close match between the
explicitly evaluated turbulent flux and the gradient-diffusion closure with the parameter
$C_{\kappa}^{(\mathrm{scl})<}=0.422$. Remarkably, the implied
turbulent kinetic Prandtl number is of the order of ten rather than
unity.

It appears that the gradient-diffusion closure provides a diffusive
mechanism which accounts for the intensity of turbulent transport but
fails to reproduce anisotropic properties of third-order generalised
moment.  This is why advanced statistical theories of turbulence
abandon the gradient-diffusion closure and introduce dynamical
equations for the third-order moments or make use of other, more
sophisticated closures \citep{Canuto97,CanDub98}. Such equations have
been suggested for the application in SGS models as well
\citep{Canuto94}.  On account of the difficulties solving these
equations, however, we prefer the simple algebraic
closure~(\ref{eq:diff_close}) with a constant diffusivity parameter
\begin{equation}
	C_{\kappa} \approx 0.4
\end{equation}
corresponding to the turbulent diffusivity
\begin{equation}
	\label{eq:diff_sgs}
	\kappa_{\mathrm{sgs}} =
	0.4\rho\Delta_{\mathrm{eff}}k_{\mathrm{sgs}}^{1/2}.
\end{equation}
According to our numerical investigation, $C_{\kappa}\approx 0.4$ is
representative for stationary isotropic turbulence of Mach number $\la
1$. In the case of developing turbulence, the effects of turbulent
transport are rather marginal, and the deviations introduced by the
statistical diffusivity~(\ref{eq:diff_sgs}) are not overly important
for the subgrid scale dynamics. For higher Mach numbers, however,
there appears to be a trend towards systematically larger
diffusivity.

In a similar fashion as the gradient-diffusion hypothesis, a turbulent
conductivity $\chi_{\mathrm{sgs}}$ for the generalised conductive flux
in fluid of heat capacity $c_{\mathrm{P}}$ and thermal conductivity
$\chi$ can be introduced:
\begin{equation}
	\vec{\mathcal{F}}^{(\mathrm{cond})}\circeq
	\rho c_{\mathrm{P}}(\chi+\chi_{\mathrm{sgs}})\vec{\nabla}T.
\end{equation}
For the generalised convective flux
$\vec{\mathcal{F}}^{(\mathrm{cond})}$, a closure might be based upon
the super-adiabatic gradient \citep{Canuto94}. Moreover, in some combustion
problems or in simulations of multi-phase media the turbulent mixing
of particle species is yet another challenge. These problems are
left for future work.

\section{Turbulent burning in a box}
\label{sc:burn_box}

As a simple testing scenario, we performed LES of turbulent
thermonuclear deflagration in degenerate carbon and oxygen.  In these
simulations, we utilised a greatly simplified reaction scheme, where
the products of thermonuclear fusion are nickel and alpha
particles. The thermonuclear burning zones propagate in a fashion
similar to premixed chemical flames. For the chosen mass density,
$\rho_{0}\approx 2.9\cdot10^{8}\,\mathrm{g\,cm^{-3}}$, the width of
the flames is $\delta_{\mathrm{F}}\approx 0.006\,\mathrm{cm}$
\citep[cf.][]{TimWoos92}.  Hence, the flame fronts are appropriately
represented by discontinuities for the numerical resolution
$\Delta=2\cdot10^{3}\,\mathrm{cm}$ in the simulations we run. The
front propagation is numerically implemented by means of the
\emph{level set method} \citep{OshSeth88,ReinHille99a}. The domain of
the flow is cubic with periodic boundary conditions (BCs). In this
scenario, the burning process consumes all nuclear fuel within finite
time.  We set $X=216\Delta=4.32\,\mathrm{km}$ for the size of the
domain, which is comparable to the resolution of the large scale
supernova simulations to be discussed in paper II.  Since self-gravity is
insignificant on length scales of the order of a few kilometres, we
apply an external solenoidal force field in order to produce turbulent
flow. Each Fourier mode of the force field is evolved as a distinct
stochastic process of the Ornstein-Uhlenbeck type.  The characteristic
wavelength $L$ of the forcing modes is half the size of the
domain. $L$ can be interpreted as integral length scale of the
flow. An detailed description of the methodology and a discussion of
numerous simulations is given in \citet{SchmHille05b}.

The LES of turbulent combustion is a particularly appropriate case
study for the performance of subgrid scale models because the
evolution of the system is strongly coupled to the SGS turbulence
energy via the turbulent flame speed relation.  For the notion of a
turbulent flame speed see \citet{Poch94}, \citet{NieHille95} and
\citet{Peters99}.  In the framework of the filtering formalism, the
underlying hypothesis is the following: If the flow is smoothed on a
certain length scale $\Delta$, then the effective propagation speed
$s_{\mathrm{turb}}(\Delta)$ of a burning front is of the order of the
turbulent velocity fluctuations $v^{\,\prime}\sim
k_{\mathrm{turb}}^{1/2}$, provided that $\Delta\gg l_{\mathrm{G}}$.
The length scale $l_{\mathrm{G}}$ is called the \emph{Gibson scale}.
It specifies the minimal size of turbulent eddies affecting the flame
front propagation. In the context of a LES, we have
$s_{\mathrm{turb}}\sim q_{\mathrm{sgs}}$ for the turbulent flame
speed.  Consequently, the SGS model determines the propagation speed
of turbulent flames. If $l_{\mathrm{G}}\lesssim\Delta$, on the other hand, the front
propagation is determined by the microscopic conductivity of the
fuel. The corresponding propagation speed is called the laminar flame
speed and is denoted by $s_{\mathrm{lam}}$.  
Since $s_{\mathrm{lam}}$ is determined by the balance between
thermal conduction and thermonuclear heat generation, conduction
effects are implicitly treated by the level set method. For this
reason, we do not include the conduction terms in
equation~(\ref{eq:energy_tot}) for the total energy.

Both limiting cases of turbulent and laminar burning, respectively,
are accommodated in the flame speed relation proposed by
\citet{Poch94}:
\begin{equation}
  \label{eq:sgs_flame_speed_pocheau}
  \frac{s_{\mathrm{turb}}}{s_{\mathrm{lam}}} = 
  \left[1 + C_{\mathrm{t}}
            \left(\frac{q_{\mathrm{sgs}}}{s_{\mathrm{lam}}}\right)^{2}
  \right]^{1/2}.
\end{equation}
The coefficient $C_{\mathrm{t}}$ is of the order unity and determines
the ratio of $s_{\mathrm{turb}}$ and $q_{\mathrm{sgs}}$ in the
turbulent burning regime.  \citet{Peters99} proposes
$C_{\mathrm{t}}=4/3$, while $C_{\mathrm{t}}\approx20/3$ is suggested
by \citet{KimMen99}.  Here we set $C_{\mathrm{t}}=1$ corresponding to
the asymptotic relation $s_{\mathrm{turb}}\simeq q_{\mathrm{sgs}}$
assumed by \cite{NieHille95}.  For a study of the influence of
$C_{\mathrm{t}}$ see \citet{SchmHille05b}.  The laminar flame speed
for an initial mass density of $2.90\cdot10^{8}\,\mathrm{g\,cm^{-3}}$
is $s_{\mathrm{lam}}\approx 9.78\cdot 10^{5}\,\mathrm{cm\,s^{-1}}$.
Choosing a characteristic velocity $V=L/T=100s_{\mathrm{lam}}$, where
$T$ is the autocorrelation time of the stochastic force driving the
flow, the Gibson scale becomes $l_{\mathrm{G}}\sim 10^{-6}L\sim
0.1\,\mathrm{cm}$. Note that the Gibson length is still large compared
to the flame thickness. Therefore, the internal structure of the
burning zones is not disturbed by turbulent velocity fluctuations,
i.e. the \emph{flamelet regime} of turbulent combustion applies
\citep{Peters99}.

\begin{figure*}[thb]
  \begin{center}
    \mbox{\subfigure[Thermonuclear burning]{\includegraphics[width=8.5cm]{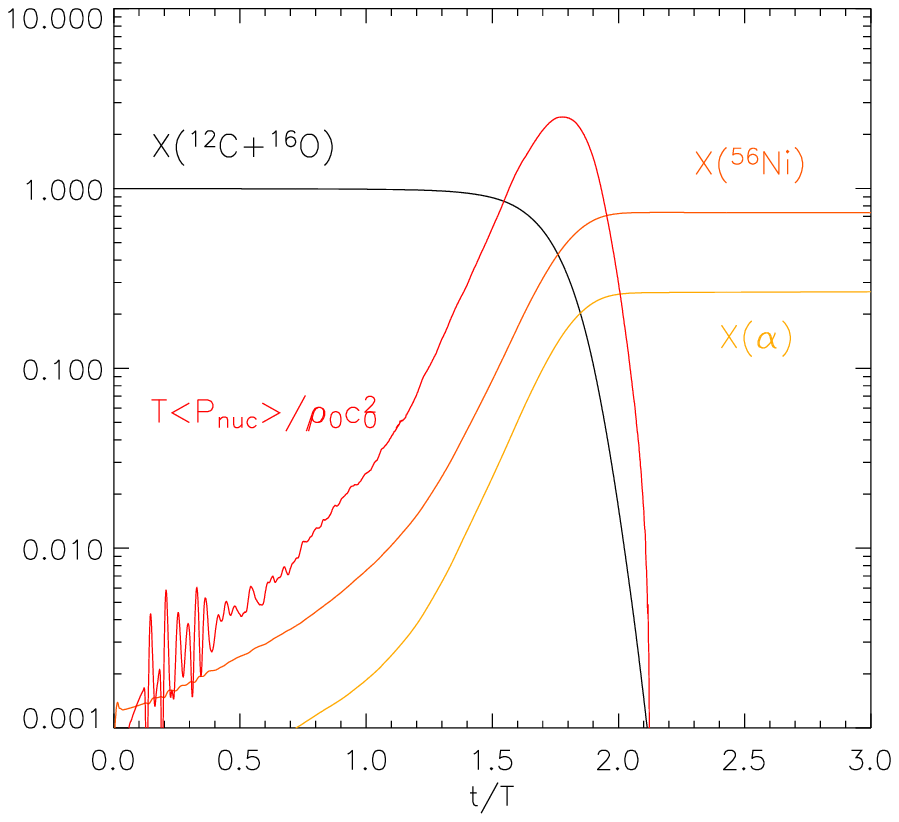}}
          \subfigure[SGS turbulence]{\includegraphics[width=8.5cm]{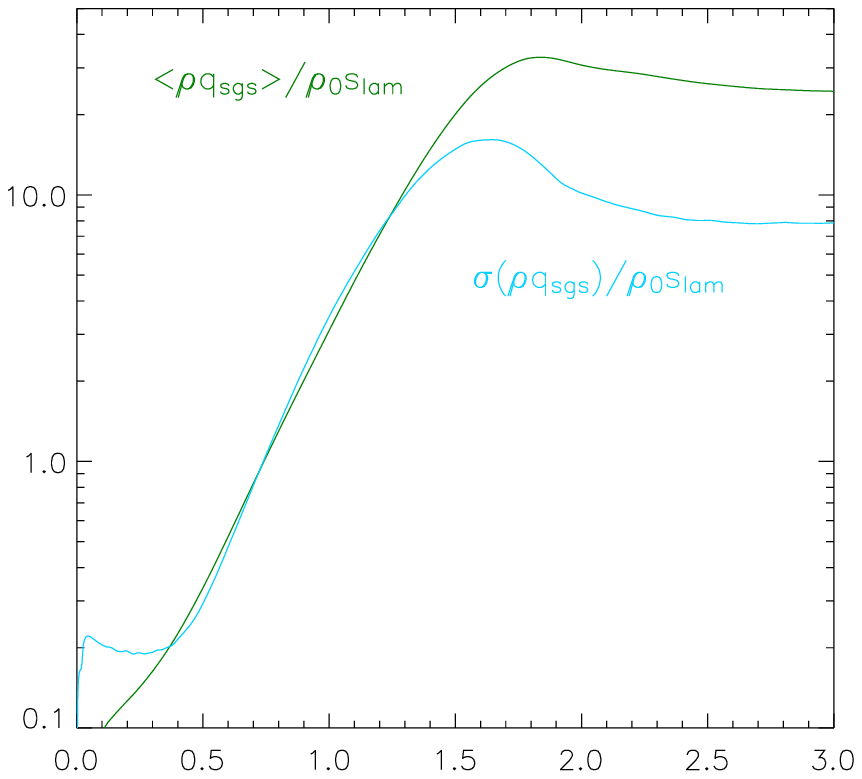}}}
    \caption{Evolution of statistical quantities in a LES of
    thermonuclear deflagration in a cubic domain subject to periodic
    boundary conditions with $N=216^{3}$ numerical cells. In panel (a)
    the spatially averaged rate of nuclear energy generation in
    combination with the mass fractions of unprocessed material
    (carbon and oxygen), alpha particles and nickel are plotted. The
    mean as well as the standard deviation of the SGS turbulence
    velocity $q_{\mathrm{sgs}}$ are shown in panel (b).}
    \label{fg:evol_burn}
  \end{center}
\end{figure*}

\begin{figure*}[thb]
  \begin{center}
    \vspace{80mm}
    \mbox{\subfigure[$t=0.25T$]{\texttt{\hspace{30mm}fig\_burn01.png\hspace{30mm}}}\qquad
          \subfigure[$t=0.50T$]{\texttt{\hspace{30mm}fig\_burn02.png\hspace{30mm}}} }\\
    \vspace{80mm}
    \mbox{\subfigure[$t=0.75T$]{\texttt{\hspace{30mm}fig\_burn03.png\hspace{30mm}}}\qquad
          \subfigure[$t=1.00T$]{\texttt{\hspace{30mm}fig\_burn04.png\hspace{30mm}}} }
    \caption{LES of thermonuclear deflagration in a cubic domain with
    periodic BCs. Shown are snapshots of the flame fronts with contour sections of
    the SGS turbulence velocity in logarithmic scaling.}
    \label{fg:burn_box1}
  \end{center}
\end{figure*}

\begin{figure*}[thb]
  \begin{center}
    \vspace{80mm}
    \mbox{\subfigure[$t=1.25T$]{\texttt{\hspace{30mm}fig\_burn05.png\hspace{30mm}}}\qquad
          \subfigure[$t=1.50T$]{\texttt{\hspace{30mm}fig\_burn06.png\hspace{30mm}}} }\\
    \vspace{80mm}
    \mbox{\subfigure[$t=1.75T$]{\texttt{\hspace{30mm}fig\_burn07.png\hspace{30mm}}}\qquad
          \subfigure[$t=2.00T$]{\texttt{\hspace{30mm}fig\_burn08.png\hspace{30mm}}} }
    \caption{Fig.~\ref{fg:burn_box1} continued.}
    \label{fg:burn_box2}
  \end{center}
\end{figure*}

\begin{figure*}[thb]
  \begin{center}
    \mbox{\subfigure[Rate of burning]{\includegraphics[width=8.5cm]{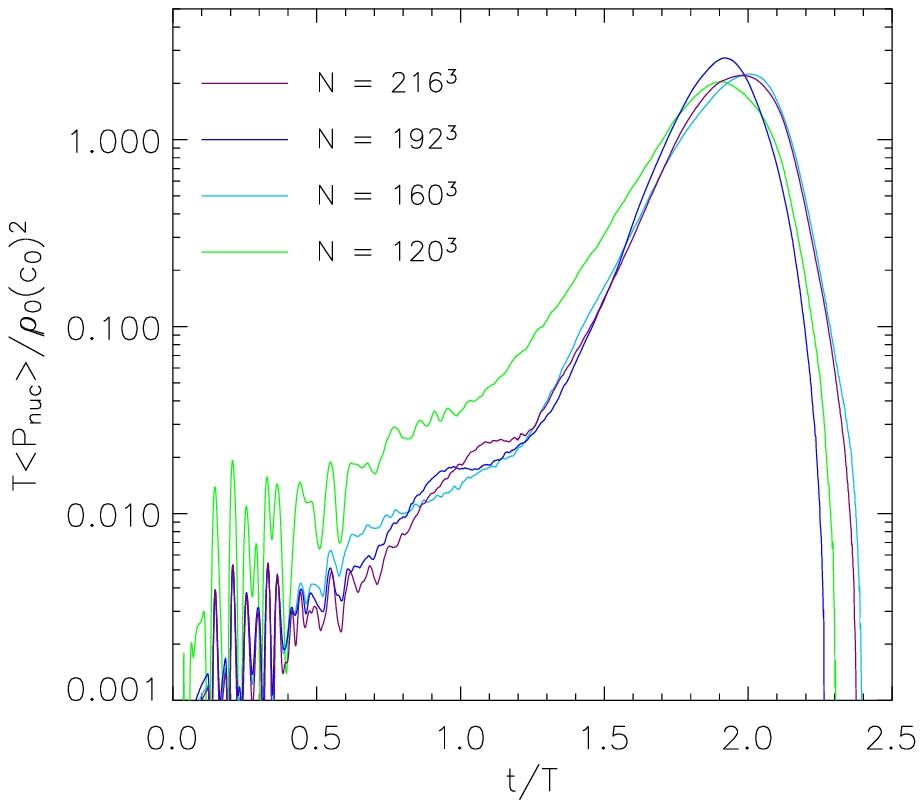}}
          \subfigure[SGS turbulence]{\includegraphics[width=8.5cm]{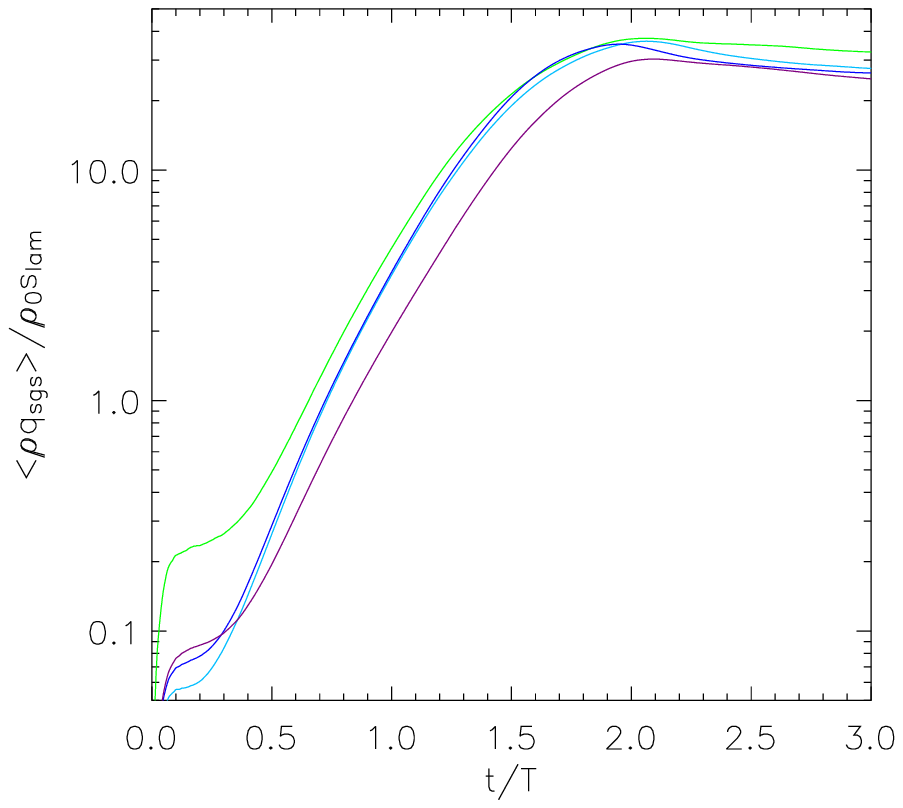}}}
    \caption{Evolution of the mean dimensionless rate of
	nuclear energy generation (a) and the ratio of the
	mass-weighted mean SGS turbulence velocity to laminar burning
	speed (b) in a sequence of LES with varying resolution. }
    \label{fg:les_res_burn}
  \end{center}
\end{figure*}

Running a LES with the parameters outlined above and setting eight
small ignition spots on a numerical grid
of $N=216^{3}$ cells, the expectation was that the burning process
would initially proceed slowly, but as turbulence was developing due
to the action of the driving force, $q_{\mathrm{sgs}}$ would
eventually exceed the laminar flame speed and substantially accelerate
the flame propagation.
Indeed, this is what can be seen in Fig.~\ref{fg:evol_burn}
which shows plots of statistical quantities as functions of time.  The
corresponding flame evolution is illustrated in the sequence of
three-dimensional visualisations in Fig.~\ref{fg:burn_box1}
and~\ref{fg:burn_box2}, where the colour shading indicates the contour
sections of $q_{\mathrm{sgs}}$ in logarithmic scaling. Initially, the
spherical blobs of burning material are expanding slowly and become
gradually elongated and folded by the onsetting flow which is produced
by the driving force.  As the SGS turbulence velocity
$q_{\mathrm{sgs}}$ exceeds the laminar burning speed
$s_{\mathrm{lam}}$ in an increasing volume of space, the spatially
averaged rate of nuclear energy release, $\langle
P_{\mathrm{nuc}}\rangle$, is increasing rapidly
(Fig.~\ref{fg:evol_burn}). Eventually, $\langle
P_{\mathrm{nuc}}\rangle$ assumes a peak value at dimensionless time
$\tilde{t}=t/T\approx 1.8$ which coincides with the maximum of
turbulence energy. Subsequently, the flow approaches statistical
equilibrium between mechanical production and dissipation of kinetic
energy. Thus, the greater part of the fuel is burned within one
large-eddy turn-over time of the turbulent flow. This observation in
combination with the tight correlation between the growth of the mean
rate of nuclear energy release and the SGS turbulence velocity
verifies that the burning process is dominated by turbulence.

As a further indicator for the reliability of the SGS model,
we varied the resolution in a sequence of LES, while maintaining the
physical parameters unaltered. The resulting global statistics is
shown in Fig.~\ref{fg:les_res_burn}. In particular, the time evolution
of $\langle P_{\mathrm{nuc}}\rangle$ appears
to be quite robust with respect to the numerical resolution.  The
deviations which can be discerned in the height, width and location of
the peak are mostly a consequence of the different flow realisations
due to the random nature of the driving force. Actually, even if we
had used identical sequences of random numbers to compute the
stochastic force field in each simulation, the dependence of the time
steps on the numerical resolution nevertheless would have produced different
discrete realisations. Thus, we initialised the random number
sequences differently and restricted the resolution study to
statistical comparisons. The evolution of the mass-weighted SGS
turbulence velocity which is plotted in panel (b) of
Fig.~\ref{fg:les_res_burn} reveals that turbulence is developing
slightly faster in the case $N=192^{3}$. This can be attributed to a
somewhat larger root mean square force field during the first
large-eddy turn-over in this simulation. Consequently, the burning
process proceeds systematically faster. Note, however, that the
level of SGS turbulence becomes monotonically lower with increasing
resolution for the almost stationary flow at time $\tilde{t}=3.0$.The
deviations for the LES with the lowest resolution ($N=120^{3}$), on
the other hand, are likely to be spurious. For this reason, it would
appear that the minimal resolution for sufficient convergence has to
be set in between $N=120^{3}$ and $N=160^{3}$.

\begin{figure}[htb]
  \begin{center}
    \resizebox{\hsize}{!}{\includegraphics[width=17cm]{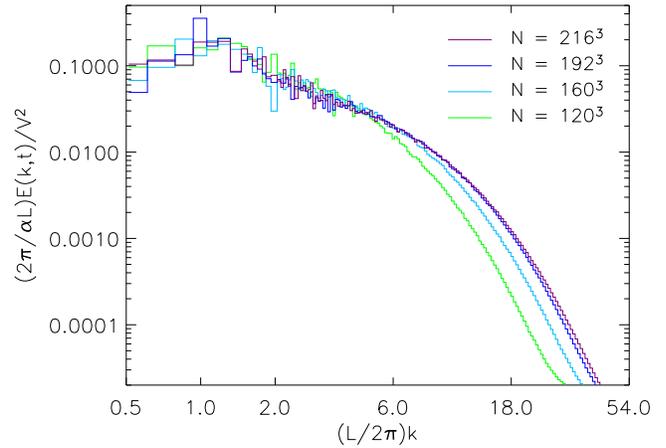}}
    \caption{Transversal kinetic energy spectrum functions at
	time $t=2T$ for the same sequence of LES as in
	Fig~\ref{fg:les_res_burn}. }
    \label{fg:les_spect}
  \end{center}
\end{figure}

This conclusion is also supported by the turbulence energy
spectra plotted in Fig.~\ref{fg:les_spect}.  We computed the
normalised energy spectrum functions for the transversal modes of the
velocity fields after two integral time scales have elapsed.  Details
of the computation of discrete spectrum functions are discussed in
\citet{SchmHille05a}.  One can clearly discern maxima in the vicinity
of the normalised characteristic wave number $\tilde{k}=Lk/2\pi=1.0$
of the driving force. For the LES with $N$ greater than $120^{3}$, an
inertial subrange emerges in the interval $2\lesssim\tilde{k}\lesssim
6$.  The dimensionless cutoff wave number in the case $N=216^{3}$ is
$\tilde{k}=54$.  As demonstrated in \citet{SchmHille05a}, the
numerical dissipation of PPM, which was used to solve the
hydrodynamical equations, noticeably smoothes the flow for wavenumbers
$\tilde{k}\gtrsim 54/9=6$. This is exactly what is observed in
Fig.~\ref{fg:les_spect}.  For $N=120^{3}$, on the other hand,
virtually all wavenumbers not directly affected by stochastic forcing
are subject to numerical dissipation, i.e.  there is no inertial
subrange at all. Considering the more common power-of-two numbers of
cells, a grid of $N=128^{3}$ cells will provide only marginally
sufficient resolution, whereas one will be on the safe side with
$N=256^{3}$.  In paper II, however, it is shown that still higher
resolutions might be required for LES of non-stationary inhomogeneous
turbulence such as in the case of thermonuclear supernova
simulations.

It is also argued in \citet{SchmHille05a} that the intrinsic
mean rate of dissipation produced by PPM closely agrees with the
prediction of the Smagorinsky model for stationary isotropic
turbulence.  This suggests that the numerical dissipation can be
utilised as an implicit SGS model with regard to the velocity
field. In fact, the LES presented in Fig.~\ref{fg:evol_burn},
\ref{fg:burn_box1} and~\ref{fg:burn_box2} was computed without
including the SGS stress term in the dynamical
equation~(\ref{eq:vel}), while the total energy $e_{\mathrm{tot}}$,
which is conserved by PPM, was coupled to the SGS turbulence energy
$k_{\mathrm{sgs}}$.  One can think of $k_{\mathrm{sgs}}$ as a buffer
between the resolved kinetic energy $\frac{1}{2}|\vec{v}|^{2}$ and the
internal energy $e_{\mathrm{int}}$. Apart from the energy budget, the
SGS model influences the resolved dynamics via the turbulent flame
speed. For the LES with varying resolution (Fig.~\ref{fg:les_res_burn}
and~\ref{fg:les_spect}), on the other hand, we applied complete
coupling of the SGS model, i.e. the turbulent stress term in the
momentum equation was included as well.  Comparing
Fig.~\ref{fg:evol_burn} (a) and~\ref{fg:les_res_burn} (a) for
$N=216^{3}$, it appears that the burning process is slightly delayed
in the latter case. As is discussed at length in \citet{SchmHille05b},
the discrepancy can be attributed to a difficulty related to inverse
energy transfer. Since backscattering injects energy on the smallest
resolved scales, which are sizeably affected by numerical dissipation,
the kinetic energy added to the flow is more or less instantaneously
converted into internal energy.  Thus, the backscattering of energy
from subgrid scales to the resolved flow results in an artificially
enhanced dissipation which depletes turbulence energy. Using partial
coupling, this unwanted effect is simply ignored. For consistency, one
must then introduce a cutoff for the eddy-viscosity parameter
$C_{\nu}$ in order to dispose of negative viscosities. Mending the
shortcoming of the treatment of inverse energy transfer is the subject
of ongoing research.  For the time being, the partial coupling of the
SGS model with backscattering suppressed serves as a pragmatic
solution in hydrodynamical simulations with PPM.

\section{Conclusion}

The localised SGS turbulence energy model offers robustness and
flexibility at relatively low computational cost. For this reason, it
is particularly suitable for the application in LES of astrophysical
fluid dynamics.  The energy transfer from resolved toward subgrid
scales is modelled with the standard eddy-viscosity closure, where the
closure parameter is computed from local properties of the
flow. Hence, there are no \emph{a priori} assumption about the resolved flow
incorporated in the model. Non-local transport is treated with the
down-gradient closure, using a constant statistical parameter.  With a
turbulent kinetic Prandtl number significantly larger than unity, it
is possible to reproduce the magnitude of diffusive flux quite
well. The rate of viscous dissipation appears to be particularly
challenging.  We found that a semi-statistical approach yields
satisfactory results.

The SGS model was implemented in a code for the LES of
turbulent thermonuclear combustion in a periodic box using the
piece-wise parabolic method (PPM) for the resolved hydrodynamics and
the level set method for the flame front propagation. Since PPM
produces significant numerical dissipation, we found it favourable to
decouple the SGS model form the momentum equation and suppressing
inverse energy transfer from unresolved toward resolved scales.  In
this kind of application, the SGS turbulence energy serves as a buffer
between the resolved kinetic energy and the internal energy and
supplies a velocity scale for calculation of the turbulent burning
speed.

Furthermore, gravitational and thermal effects can be included in the
SGS model, although closures specific to a certain physical system have
to be formulated.  An example is presented in paper II, where the
application of the SGS model to Rayleigh-Taylor-driven thermonuclear
combustion in type Ia supernova is discussed. Adapting the model to
other applications, possibly with different numerical techniques, is
the goal of on-going research.

\begin{acknowledgements}
The simulations of forced isotropic turbulence were run on the Hitachi
SR-8000 of the Leibniz Computing Centre. For the LES of turbulent
combustion we used the IBM p690 of the Computing Centre of the
Max-Planck-Society in Garching, Germany. The research of W. Schmidt
and J. C. Niemeyer was supported by the Alfried Krupp Prize for Young
University Teachers of the Alfried Krupp von Bohlen und Halbach
Foundation.
\end{acknowledgements}

\bibliographystyle{aa}

\bibliography{3617}

\end{document}